\documentclass[a4paper,11pt]{article}

\usepackage{jheppub}
\usepackage[english]{babel}
\usepackage{relsize}
\usepackage{mathrsfs}  

\usepackage{amstext}  
\usepackage{amsfonts}
\usepackage{amsthm}
\usepackage{graphicx}
\usepackage{amsmath}
\usepackage{amsfonts}
\usepackage{mathtools}

\usepackage{array}                
\usepackage{xcolor} 

\usepackage{slashed}

\newcommand{\OpenLoops}{{\rmfamily\scshape OpenLoops}}
\newcommand{\OpenLoopstwo}{{\rmfamily\scshape OpenLoops\,2}}
\newcommand{\Collier}{{\rmfamily\scshape Collier}}

\renewcommand{\refeq}[1]{\mbox{\eqref{#1}}}

\newcommand{\reffi}[1]{\mbox{Fig.~\ref{#1}}}

\newcommand{\refta}[1]{\mbox{Table~\ref{#1}}}

\newcommand{\refse}[1]{\mbox{Section~\ref{#1}}}
\newcommand{\refses}[2]{\mbox{Sections~\ref{#1}--\ref{#2}}}
\newcommand{\refapp}[1]{\mbox{Appendix~\ref{#1}}}

\newcommand{\ie}{i.e.\ }

\newcommand{\f}[2]{\frac{#1}{#2}}
\newcommand{\sss}[1]{\scriptscriptstyle#1}
\newcommand{\ssst}[1]{\scriptscriptstyle{\text{#1}}}
\newcommand{\nosss}[1]{#1}

\newcommand{\bea}{\begin{eqnarray}}
\newcommand{\eea}{\end{eqnarray}}
\newcommand{\be}{\begin{equation}}
\newcommand{\ee}{\end{equation}}
\newcommand{\ba}{\begin{align}}
\newcommand{\ea}{\end{align}}
\newcommand{\beas}{\begin{eqnarray*}}
\newcommand{\eeas}{\end{eqnarray*}}
\newcommand{\bes}{\begin{equation*}}
\newcommand{\ees}{\end{equation*}}
\newcommand{\bas}{\begin{align*}}
\newcommand{\eas}{\end{align*}}

\newcommand{\eps}{{\varepsilon}}

\newcommand{\lb}{\left(}
\newcommand{\rb}{\right)}

\newcommand{\idop}{1\!\!1}

\newcommand{\Db}[2]{{D}_{\nosss{#1}}^{(#2)}}
\newcommand{\Dbi}[1]{{D}_{\nosss{#1}}^{(1)}}

\newcommand{\momk}[1]{k_{\nosss{#1}}}
\newcommand{\momp}[1]{p_{#1}}
\newcommand{\mass}[1]{m_{\nosss{#1}}}
\newcommand{\heli}{h}
\newcommand{\helibar}{\bar h}
\newcommand{\helicheck}{\check h}

\newcommand{\helihat}{\hat h}

\newcommand{\momq}{\bar{q}}

\newcommand{\npart}{N}
\newcommand{\calA}{\mathcal{A}}
\newcommand{\calC}{\mathcal{C}}

\newcommand{\calK}{\mathcal{K}}
\newcommand{\calM}{\mathcal{M}}
\newcommand{\calMCT}{\mathcal{M}^{\ssst{(CT)}}}
\newcommand{\calN}{\mathcal{N}}
\newcommand{\barN}{\bar{\mathcal{N}}}

\newcommand{\barM}{\bar{\mathcal{M}}}
\newcommand{\calMstar}[1]{\calM^{\hspace{.3ex}\displaystyle *}_{#1}}
\newcommand{\barMstar}[1]{\barM^{\hspace{.3ex}\displaystyle *}_{#1}}

\newcommand{\calE}{\mathcal{E}}
\newcommand{\calV}{\mathcal{V}}
\newcommand{\calW}{\mathcal{W}}
\newcommand{\seg}{S}

\newcommand{\nhcs}{N_{\mathrm{hcs}}}

\newcommand{\calU}{\mathcal{U}}
\newcommand{\helicheckloc}{\tilde h}

\newcommand{\segment}[2]{\seg_{#2}^{(#1)}}

\newcommand{\subtree}[2]{w_{#2}^{(#1)}}
\newcommand{\particlesetsegment}[2]{\calE_{#2}^{(#1)}}

\newcommand{\helisegment}[2]{\heli_{#2}^{(#1)}}

\newcommand{\helic}[1]{\heli^{(#1)}}                   
\newcommand{\helig}{\heli}                             
\newcommand{\helicc}[1]{\helicheck^{(#1)}}             
\newcommand{\helipc}[2]{\helihat_{#2}^{(#1)}}          
\newcommand{\helipcc}[2]{\helicheck_{#2}^{(#1)}}       
\newcommand{\helipccloc}[2]{\helicheckloc_{#2}^{(#1)}} 

\newcommand{\numc}[1]{\calN^{(#1)}}                 
\newcommand{\numpc}[2]{\calN^{(#1)}_{#2}}           
\newcommand{\numi}{\calU^{(1)}}                 
\newcommand{\numpi}[1]{\calU^{(1)}_{#1}}           
\newcommand{\numpii}[1]{\calU^{(2)}_{#1}}           
\newcommand{\numpd}[1]{\calU^{(123)}_{#1}}           
\newcommand{\numinter}[1]{{\calU_{#1}^{(13)}}}                

\newcommand{\indc}[2]{\beta^{\sss{(#1)}}_{\sss{#2}}}
\newcommand{\indci}[1]{\beta^{\sss{(1)}}_{\sss{#1}}}
\newcommand{\indcii}[1]{\beta^{\sss{(2)}}_{\sss{#1}}}
\newcommand{\indciii}[1]{\beta^{\sss{(3)}}_{\sss{#1}}}

\newcommand{\amp}[2]{{\calA}_{{#1},{#2}}}

\newcommand{\fullamp}[2]{{\calM}_{{#1},{#2}}}

\newcommand{\fullampbar}[2]{{\barM}_{{#1},{#2}}}
\newcommand{\colfac}[2]{{C}_{{#1},{#2}}}
\newcommand{\vertex}[1]{\calV_{#1}}
\newcommand{\calCh}[1]{\mathcal{C}_{#1}}  
\newcommand{\denc}[1]{\mathcal{D}^{(#1)}(\bar q_{#1})} 
\newcommand{\dencexpl}[2]{\mathcal{D}^{(#1)}( #2)} 
\newcommand{\col}{\mathrm{col}}
\newcommand{\hel}{\mathrm{hel}}
\newcommand{\re}{\mathrm{Re}}
\newcommand{\Tr}{\mathrm{Tr}}

\newcommand{\redone}{Red1}
\newcommand{\redtwo}{Red2}

\newcommand{\rB}{\mathrm B}

\newcommand{\ri}{\mathrm i}
\newcommand{\rd}{\mathrm d}
\newcommand{\ord}{\mathcal O}

\newcommand{\LO}{\mathrm{LO}}
\newcommand{\NLO}{\mathrm{NLO}}
\newcommand{\NNLO}{\mathrm{NNLO}}

\newcommand{\tVV}{t_{\mathrm{VV}}}
\newcommand{\tRV}{t_{\mathrm{RV}}}
\newcommand{\tRVfull}{t_{\mathrm{RV}}^{\ssst{full}}}
\newcommand{\tRRV}{t_{\mathrm{RRV}}}
\newcommand{\tRRVfull}{t_{\mathrm{RRV}}^{\ssst{full}}}
\newcommand{\Ndiags}{N_{\mathrm{diags}}}

\newcommand{\Nhel}[1]{N^{(#1)}_{\hel}}
\newcommand{\Ncoeff}[1]{N^{(#1)}_{\mathrm{tens}}}

\definecolor{bluemar}{rgb}{0,0,.5}
\definecolor{redmar}{rgb}{.8,0,0}
\definecolor{greenmar}{rgb}{0,.5,0}

\graphicspath{{./pdfdiagrams/}}

\preprint{
\begin{flushright}
PSI PR-22-03\\
ZU-TH 03/22\\
\end{flushright}
}

\title{\boldmath Two-loop tensor integral coefficients in OpenLoops}

\author[a]{Stefano Pozzorini}
\author[b]{Natalie Sch\"ar}
\author[b]{Max F.~Zoller}

\affiliation[a]{Physik-Institut, Universit\"at Z\"urich, Winterthurerstrasse 190, 
CH-8057 Z\"urich, Switzerland}
\affiliation[b]{Paul Scherrer Institut, Forschungsstrasse 111, CH-5232 Villigen PSI, Switzerland}

\emailAdd{pozzorin@physik.uzh.ch}
\emailAdd{natalie.schaer@psi.ch}
\emailAdd{max.zoller@psi.ch}

\abstract{We present a new and fully general algorithm for the automated
construction of the integrands of two-loop scattering amplitudes.
This is achieved through a generalisation of the open-loops method to two
loops.
The core of the algorithm consists of a numerical recursion, where the
various building blocks of two-loop diagrams are connected to each other
through process-independent operations that depend only on the Feynman rules
of the model at hand.
This recursion is implemented in terms of tensor coefficients that encode the
polynomial dependence of loop numerators on the two independent loop
momenta.
The resulting coefficients are ready to be combined with corresponding
tensor integrals to form scattering probability densities at two loops.
To optimise CPU efficiency we have compared several algorithmic options
identifying one that outperforms naive solutions by two orders of magnitude.
This new algorithm is implemented in the \OpenLoops{}
framework in a fully automated way for
two-loop QED and QCD corrections to any Standard Model process.
The technical performance is discussed in detail for several $2\to2$ and
$2\to 3$ processes with up to order $10^5$ two-loop diagrams.  We find that the 
CPU cost scales linearly with the number of two-loop diagrams
and is comparable to the cost of corresponding real--virtual ingredients in a NNLO
calculation.
This new algorithm constitutes a key building block for
the construction of an automated generator of scattering amplitudes at two
loops.
}

\arxivnumber{}

\begin{document}
\maketitle
\flushbottom

\newpage

\section{Introduction}

Precise theory predictions 
based on the Standard Model (SM) play a crucial role for
the success of the LHC
physics program. In particular, in the absence of striking signals of new physics,
the availability of high theory precision for the widest possible range of
scattering processes
is a key prerequisite in order to maximise the sensitivity 
of LHC measurements to small effects of 
physics beyond the SM.

Predictions at next-to-leading order (NLO)
in perturbation theory can be 
obtained through multi-purpose
Monte Carlo tools~\cite{Gleisberg:2008ta,
Alioli:2010xd,
Bevilacqua:2011xh,
Alwall:2014hca},
and the hard scattering amplitudes at the core of these calculations are computed with 
automated numerical tools~\cite{Ossola:2007ax,
Berger:2008sj,
vanHameren:2009dr,
vanHameren:2010cp,
Hirschi:2011pa,
Cascioli:2011va,
Badger:2012pg,
Cullen:2014yla,
Peraro:2014cba,
Denner:2016kdg,
Actis:2016mpe,
Carrazza:2016gav,
Buccioni:2019sur}
at tree and one-loop level.

At next-to-next-to-leading order (NNLO)
the required two-loop amplitudes 
still need to be derived on a process-by-process basis.
To date, full NNLO predictions for $2\to 2$ processes
are widely available (see~\cite{Boughezal:2015dra,
Heinrich:2017bvg,
Gehrmann-DeRidder:2019ibf,
deFlorian:2019app,
Grazzini:2019jkl,
Czakon:2020coa,
Banerjee:2021mty, 
Banerjee:2021qvi,
Campbell:2021mlr,
Buccioni:2022kgy} for recent progress\footnote{%
This list of references includes only a few representative 
recent results.
For a recent review see~\cite{Heinrich:2020ybq}.
}
),
while for  $2\to 3$ processes only 
a few two-loop amplitudes~\cite{Badger:2018enw,
Abreu:2018zmy,
Abreu:2019odu,
Badger:2019djh,
Hartanto:2019uvl,
Chawdhry:2020for,
Chawdhry:2021mkw, 
Agarwal:2021grm,
Agarwal:2021vdh, 
Abreu:2021oya, 
Abreu:2021asb, 
Badger:2021nhg,
Badger:2021owl, 
Badger:2021imn,
Badger:2021ega, 
Badger:2022ncb}  
and pioneering NNLO results~\cite{Kallweit:2020gcp, 
Chawdhry:2019bji,
Chawdhry:2021hkp,
Czakon:2021mjy}
exist.
The complexity and the status of NLO calculations for loop-induced processes
is similar, with an increasing number of 
$2\to 2$ predictions~\cite{Caola:2018zye,Davies:2019dfy, Grazzini:2020stb,
Bonetti:2020hqh,Becchetti:2021axs}
and first results for $2\to 3$ processes~\cite{Badger:2021ohm}.
Calculations at N$^3$LO are 
highly challenging and 
currently restricted to
\mbox{$2\to 1$} processes~\cite{%
Anastasiou:2015ema,  
Anastasiou:2016cez, 
Cieri:2018oms,
Duhr:2019kwi,        
Chen:2021isd,    
Duhr:2020sdp,    
Duhr:2021vwj,
Camarda:2021ict}.    

While specialised NNLO Monte Carlo 
tools~\cite{Grazzini:2017mhc,Gauld:2019ntd,Banerjee:2020rww} have been developed, a fully
general and automated NNLO tool, 
including in particular the
two-loop scattering amplitudes, is not yet within reach.
In light of the high-precision requirements of the LHC and future colliders,
such a NNLO tool for 
arbitrary $2\to 2$ and $2\to 3$ SM processes
is highly desirable.
With this objective in mind, 
in this paper we present a new algorithm that 
provides an important building block 
for the automated construction of two-loop amplitudes.

This new algorithm represents the extension of the well-established
\OpenLoops{} technique~\cite{Cascioli:2011va} from one to two loops. In the  
\OpenLoops{} approach, one-loop scattering amplitudes are constructed as sums of
Feynman diagrams. Individual one-loop diagrams  $\Gamma$ have the form 
\be
\fullampbar{1}{\Gamma} = 
\colfac{1}{\Gamma}
\int\!\rd\momq_1
\f{\barN(\momq_1)}
{\mathcal{D}(\momq_1)}\,,
\label{eq:intro_amp_one}
\ee
where $\colfac{1}{\Gamma}$ is a colour factor, 
and the bar indicates quantities that are defined in $D$ dimensions in order to regularise
divergences in loop integrals \cite{tHooft:1972tcz}.
The denominator $\mathcal{D}(\momq_1)$ is a product of propagator denominators
depending on the loop momentum $\momq_1$.
In \OpenLoops{} the numerator $\barN(\momq_1)$ is
decomposed into loop-momentum tensors,
\be
\barN(\momq_1) = \sum\limits_{r}
\barN_{\mu_1\ldots\mu_r}
\momq_1^{\mu_1}\ldots \momq_1^{\mu_r}\,,
\label{eq:intro_tdec_one}
\ee
and the tensor coefficients $\barN_{\mu_1\ldots\mu_r}$
are constructed through a 
numerical recursion~\cite{vanHameren:2009vq,Cascioli:2011va,Buccioni:2017yxi}
based on process-independent
operations, which depend only on the Feynman rules of the model at hand.
This construction is carried out in four space-time dimensions, and the
contribution of the missing numerator parts of order $(D-4)$ 
are reconstructed via insertion of process-independent rational
counterterms~\cite{Ossola:2008xq,Draggiotis:2009yb,Garzelli:2009is,Pittau:2011qp}
into tree diagrams.
Scattering amplitudes are obtained by combining the tensor coefficients with 
the associated tensor integrals
\be
I^{\mu_1\ldots\mu_r}
\,=\, 
\int\!\rd\momq_1
\frac{q_1^{\mu_1}\ldots q_1^{\mu_r}}
{\mathcal{D}(\momq_1)}\,,
\ee
which can be reduced to scalar integrals using external
tools~\cite{Denner:2016kdg,Ossola:2007ax}.
Alternatively, the construction of tensor coefficients and the 
reduction of tensor integrals can be combined in a single numerical
recursion using the on-the-fly reduction method~\cite{Buccioni:2017yxi}.

Besides all required scattering amplitudes for NLO calculations,
the public \OpenLoops{} program provides also some of the
building blocks of NNLO calculations, namely squared one-loop amplitudes
as well as Born one-loop interferences
for processes with one unresolved emission, 
which enter 
the  real--virtual parts of NNLO cross sections.

In this paper we present a new algorithm for 
the construction of the interference of two-loop and Born amplitudes,
which enters the virtual--virtual parts of NNLO calculations.
Also in this case, full scattering processes are handled as sums 
over individual Feynman diagrams. 
The amplitude of a generic two-loop diagram $\Gamma$ has the form
\be
\fullampbar{2}{\Gamma} = 
\colfac{2}{\Gamma}
\int\!\rd\momq_1\int\!\rd\momq_2 
\f{\barN(\momq_1,\momq_2)}
{\mathcal{D}(\momq_1,\momq_2)} \label{eq:intro_amp_two}\,,
\ee
where $\colfac{2}{\Gamma}$ is a colour factor, 
and the denominator
$\mathcal{D}(\momq_1,\momq_2)$ embodies all propagator denominators
depending on the two independent loop momenta $\momq_1, \momq_2$. 
Similarly as in~\eqref{eq:intro_tdec_one},
for the numerator we apply a tensor decomposition in the two loop momenta,
\be
\barN(\momq_1,\momq_2) =
\sum\limits_{r,s}\barN_{\mu_1\ldots\mu_r,\,\nu_1\ldots\nu_s}\,
\momq_1^{\mu_1}\ldots \momq_1^{\mu_r} \momq_2^{\nu_1}\ldots
\momq_2^{\nu_s}\,.\label{eq:intro_tdec_two}
\ee
The main challenges in the development of a fully automated two-loop tool
are the efficient construction of the tensor coefficients
$\barN_{\mu_1\ldots\mu_r,\,\nu_1\ldots\nu_s}$,
the reduction of the 
associated two-loop tensor integrals to a set of
master integrals, 
and the evaluation of these master integrals.

At higher loop levels, a fully automated tensor-integral library in the
style of the existing one-loop tools
\cite{Denner:2016kdg,vanHameren:2010cp,Ossola:2007ax} is not yet available. 
However, concerning the integral reduction,
the tools and techniques based on integration-by-parts relations~\cite{ChetyrkinMINCER}
have been greatly advanced in the recent years~\cite{vonManteuffel:2012np,vonManteuffel:2014ixa, Chawdhry:2018awn,
Smirnov:2019qkx,Klappert:2020nbg}.
Furthermore, differential equations methods~\cite{Gehrmann:1999as,Henn:2013pwa} for 
the computation of Feynman integrals have greatly expanded the range of available two-loop master
integrals (see e.g.~\cite{Papadopoulos:2015jft,
Gehrmann:2018yef,
Chicherin:2018mue,
Chicherin:2020oor,
Abreu:2020jxa,
Abreu:2021smk,
Duhr:2021fhk}).

The new algorithm presented in this paper 
deals with the construction of the two-loop tensor coefficients 
$\barN_{\mu_1\ldots\mu_r,\,\nu_1\ldots\nu_s}$
in \eqref{eq:intro_tdec_two}.
This is achieved through a process-independent numerical recursion, 
which includes also the effect of the interference with the
Born amplitude as well as the summation over all colour and helicity degrees of
freedom.
This algorithm has been implemented in a fully automated way 
within the \OpenLoops{} framework for 
two-loop QCD and QED corrections to any SM process. Due to the generality of the algorithm
it can easily be extended to other models.
Similarly as in the one-loop case, all tensor coefficients are constructed in 
four space-time dimensions, and the contributions associated with the 
$\ord(D-4)$ parts of loop numerators can be reconstructed 
by means of two-loop rational
counterterms~\cite{Pozzorini:2020hkx,Lang:2020nnl,Lang:2021hnw}.
More precisely, renormalised two-loop amplitudes  in $D$ dimensions 
can be obtained from two-loop amplitudes with four-dimensional loop numerators
together with insertions of one-loop and two-loop counterterms
of rational and UV kind into one-loop and tree amplitudes.
This
approach is fully established for rational terms of UV origin, while
we expect that rational terms of infrared (IR) origin 
cancel in IR-subtracted amplitudes.
This cancellation mechanism is well understood at 
one loop~\cite{Bredenstein:2008zb} and 
is presently under investigation at the 
two-loop level. In this case, we expect 
that all rational terms of IR origin can be cancelled 
by means of a simple and process-independent 
modification of the well-established procedures for the subtraction of two-loop IR divergences.

The paper is organised as follows.
In \refse{sec:scattampNNLO} we outline general aspects of the
construction of scattering amplitudes and 
cross sections at NLO and NNLO.
In this context we discuss rational counterterms, 
as well as the bookkeeping of colour and helicity 
degrees of freedom.
In \refse{sec:treeandoneloopamp} we review 
the \OpenLoops{} method for the construction of 
tree and one-loop amplitudes. 
This sets the stage for the discussion of the new two-loop 
algorithm.
\refse{sec:twoloopred} deals with reducible two-loop diagrams, which 
can be decomposed into one-loop subdiagrams. For this kind of 
two-loop diagrams we present an algorithm that exploits and extends 
various features of the \OpenLoops{} method at one loop.
In \refse{sec:twoloopirred} we present the main novelty of this paper, a
fully general numerical algorithm for the construction of the tensor
coefficients of irreducible two-loop diagrams.  We first analyse the
structure of a generic two-loop algorithm, and then present the most efficient
algorithm that we have identified through a systematic CPU cost analysis.
The implementation of this new algorithm in the \OpenLoops{} framework and
its technical performance are discussed in~\refse{eq:performance},
where we show that the CPU cost of two-loop tensor coefficients scales
linearly in the number of Feynman diagrams and is comparable to the cost of
the real--virtual building blocks in a NNLO calculation.
Our conclusions are presented in \refse{sec:conclusion}.

\section{Scattering amplitudes and partonic cross sections up to NNLO}\label{sec:scattampNNLO}

Partonic cross sections are derived from scattering amplitudes, which in perturbation theory are computed up to a fixed loop order,
\be
\barM = \barM_0 + \barM_1 + \barM_2 + \ldots {}.\label{eq:def_matel}
\ee
The bar marks quantities in $D=4-2\eps$ dimensions throughout this paper.
The $L$-loop matrix element $\barM_{L}$ is computed 
as the sum over all $L$-loop Feynman diagrams $\Gamma$ of the process at hand, 
\be
\label{eq:amplitudes}
\barM_{L}(\heli)
=
\sum\limits_{\Gamma} 
\fullampbar{L}{\Gamma}(\heli) {},
\ee
where the argument $\heli$ corresponds to the
helicity configuration of all external particles as described in \refse{sec:helbookkeeping}. Loop integrals can exhibit UV divergences, which are cancelled through the renormalisation procedure.
We denote a renormalised amplitude as 
\be
\mathbf{R}\,\barM = \barM_0 + \mathbf{R}\,\barM_1 + \mathbf{R}\,\barM_2 + \ldots {},
\ee
where the renormalisation of the fields, couplings and masses in an $L$-loop amplitude is
implemented through counterterm insertions into lower-loop amplitudes. 
In a renormalisable model, the finite set of UV counterterms is computed once and for all in the chosen renormalisation scheme.
Differential cross sections are obtained from the colour- and helicity-summed scattering probability density
\bea
\calW &=& \frac{1}{\nhcs}\sum_{\heli}\sum_{\col}\left|\mathbf{R}\barM(\heli)\right|^2{}, \label{eq:Wdef}
\eea
where the initial-state average and symmetry factors for identical final-state particles are encoded in 
\bea
\nhcs &=&
\left(\prod_{i\in\mathcal{S}_{\mathrm{in}}} 
N_{\mathrm{hel},i}
N_{\mathrm{col},i}\right)
\left(\prod_{p\in\mathcal{P}_{\mathrm{out}}} 
n_p!\right)\,.
\eea
Here $\mathcal{S}_{\mathrm{in}}$ is the set of initial-state particles, 
and $\mathcal{P}_{\mathrm{out}}$ the set of final-state particle and antiparticle types\footnote{Note that particles and antiparticles count as different types here.},
while $N_{\mathrm{hel},i}$ and  $N_{\mathrm{col},i}$ are the number of helicity and colour states of particle $i$, and $n_p$ the number of identical final-state particles of type $p$.

The probability density \eqref{eq:Wdef} is split into contributions of different orders in perturbation theory. For processes with $\barM_{0}\neq 0$ the
leading-order (LO) and
higher-order virtual contributions are
\be
\calW = \underbrace{\calW_{00}}_{\text{LO}}
\;+ \underbrace{\calW_{01}}_{\text{NLO virtual}} 
+\; \underbrace{\calW_{02} + \calW_{11}}_{\text{NNLO virtual}} 
\;+\; \ldots{}\;, \label{eq:WNNLO}
\ee
with the $L$-loop squared and the Born $L$-loop interference terms
\bea
\calW_{LL} &=& 
\frac{1}{\nhcs}\sum_{\heli}\sum_{\col} |\mathbf{R}\barM_{L}(\heli)|^2
\qquad (L = 0,1)\label{M2WvirtLL}\,, \\
\calW_{0L} &=& 
\frac{1}{\nhcs}\sum_{\heli}\sum_{\col} 2\,\re \Big[\barMstar{0}(\heli)\,
\mathbf{R}\barM_{L}(\heli)\Big]\qquad (L = 1,2)\,. \label{eq:M2WvirtBL}
\eea
 For loop-induced processes $\calW_{11}$ is the leading order contribution.

Finite partonic cross sections require, in addition to the UV renormalisation, 
the inclusion of real-emission contributions in order to cancel final-state collinear and soft divergences, as well as
the factorisation of initial-state collinear singularities, which are absorbed into the parton distribution functions.
For the amplitudes of scattering processes with $X$ additional unresolved partons,
in analogy with \eqref{eq:def_matel} and \eqref{eq:WNNLO}, we use the notation
\be
\barM^{(X)} = \barM^{(X)}_0 + \barM^{(X)}_1 + \ldots\;, \label{eq:def_matelX}
\ee
and
\be
\calW^{(X)} = \calW^{(X)}_{00} + \calW^{(X)}_{01} +\ldots{}\;. \label{eq:WNNLOX}
\ee
In addition to \eqref{eq:WNNLO}, calculations up to NNLO require
the NLO real and NNLO real--virtual contributions
\bea
\calW^{(1)}\Big|_{\NNLO} &=& 
\underbrace{\calW_{00}^{(1)}}_{\text{NLO real}} +
\underbrace{\calW_{01}^{(1)}}_{\text{NNLO real--virtual}}\,{}, \label{eq:M2WwithX1}
\eea
as well as the NNLO double-real contribution
\bea
\calW^{(2)}\Big|_{\NNLO} &=& 
\underbrace{\calW_{00}^{(2)}}_{\text{NNLO real-real}}. \label{eq:M2WwithX2}
\eea
Partonic cross sections up to NNLO are computed as
\bea
\hat\sigma_{} &=& \hat\sigma_{\LO} + \Delta\hat\sigma_{\NLO} + \Delta\hat\sigma_{\NNLO} +
\ldots\;,
\eea
with
\bea
\hat\sigma_{\LO} &=& 
\int\!\rd\Phi_N\, \calW_{00}, \label{eq:sigmapart_LO}\\
\Delta\hat\sigma_{\NLO} &=& 
\int\!\rd\Phi_N\, \calW_{01} 
+ \int\!\rd\Phi_{N+1}\,\calW_{00}^{(1)}, \label{eq:sigmapart_NLO}\\
\Delta\hat\sigma_{\NNLO} &=& 
\int\!\rd\Phi_N\, \lb\calW_{02} + \calW_{11}\rb 
+ \int\!\rd\Phi_{N+1}\,\calW_{01}^{(1)} 
+ \int\!\rd\Phi_{N+2}\,\calW_{00}^{(2)}, \label{eq:sigmapart_NNLO}
\eea
where $\rd\Phi_{N+X}$ contains the integration measure of the Lorentz-invariant phase space
with $N$ resolved and $X$ unresolved particles as well as the flux factor. In general,
only the full sum in \eqref{eq:sigmapart_NLO} and \eqref{eq:sigmapart_NNLO} is IR-finite. A number of powerful methods for the subtraction of IR divergences at the level of individual terms in \eqref{eq:sigmapart_NLO} and \eqref{eq:sigmapart_NNLO} is available
\cite{GehrmannDeRidder:2005cm,Catani:2007vq,Somogyi:2005xz,Czakon:2010td,Boughezal:2015eha,Cacciari:2015jma}.

The program \OpenLoopstwo~\cite{Buccioni:2019sur} supports the 
calculation of 
tree-level and one-loop amplitudes, i.e.~the contributions \eqref{M2WvirtLL} for
$L=0,1$ and \eqref{eq:M2WvirtBL} for $L=1$, and hence also the 
various
NLO and NNLO
contributions in \eqref{eq:M2WwithX1} and \eqref{eq:M2WwithX2},
for any process.
This is achieved through automated algorithms that 
feature high CPU efficiency 
and numerical stability.
In particular, thanks to targeted analytic expansions and
a hybrid-precision approach~\cite{Buccioni:2017yxi,Buccioni:2019sur},
\OpenLoopstwo{} guarantees a numerically stable evaluation of
the real--virtual  NNLO contributions $\calW_{01}^{(1)}$
in the full phase space, including
the regions where the unresolved radiation
becomes highly soft or collinear.
This high degree of stability in the IR regions 
was demonstrated through successful applications of 
\OpenLoopstwo{} to state-of-the-art 
NNLO calculations based on local subtraction methods,
such as in the recent 
NNLO calculations of $pp\to 3$\,jets~\cite{Czakon:2021mjy,Chen:2022ktf}.

In this paper, we present a new algorithm for the efficient numerical
computation of the Born two-loop interference $\calW_{02}$, defined in
\eqref{eq:M2WvirtBL}, at the level of tensor-integral coefficients.

\subsection{Dimensional regularisation and rational terms{}} \label{sec:rationalterms}

\def\Dn{D_{\mathrm{n}}}

In the \OpenLoops{} framework, UV and IR divergences are regularised in 
the 't~Hooft--Veltman scheme~\cite{tHooft:1972tcz}, 
where external wave functions and momenta are four-dimensional, while 
loop momenta, metric tensors and Dirac $\gamma$-matrices inside the loops live in 
$ D=4-2\eps $ dimensions.
The integration measure as well as the 
denominators of loop integrands are kept in 
$D$ dimensions throughout, while loop-integrand numerators are split into a
four-dimensional part and a remainder of $\mathcal{O}(\eps)$.  
Contributions stemming from the former part 
are referred to as $\Dn=4$ 
dimensional, where $\Dn$ denotes the dimensionality 
of the loop numerator,
and
are calculated with numerical
algorithms where the loop momenta, metric tensors and $\gamma$-matrices in
the numerator are handled in four dimensions. 
The remaining $(D-4)$-dimensional numerator parts give rise to additional 
rational contributions, which originate from 
their interplay with the divergences of loop integrals
and can be reconstructed through process-independent counterterms.

At one loop, such rational terms originate only from 
UV poles~\cite{Bredenstein:2008zb}, and the corresponding
counterterms are known for the full SM and for a variety of other 
models~\cite{Ossola:2008xq, Draggiotis:2009yb, Garzelli:2009is, Pittau:2011qp}.
Using rational counterterms,
renormalised one-loop amplitudes in $D$ dimensions 
can be obtained from quantities in 
$\Dn=4$ dimensions through the formula
\bea 
{\textbf{R}}\, \barM_{1} (\heli)
&=&  \calM_1 (\heli) + \calMCT_{0,1}(\heli)
\,.
\label{eq:masterformula1}
\eea
Here and in the following, 
loop amplitudes $\barM_L$ carrying a bar are fully $D$-dimensional as introduced
in~\refeq{eq:def_matel}, while
amplitudes $\calM_L$ without a bar 
are in $\Dn=4$ dimensions, \ie they are computed with four-dimensional
integrand numerators and $D$-dimensional denominators.
The first term on the rhs of~\refeq{eq:masterformula1}
is the unrenormalised amplitude in $\Dn=4$, and
$\calMCT_{0,1}$ stands for the tree-level amplitude 
with all relevant insertions of UV and rational 
one-loop counterterms.

At two loops, as recently shown in~\cite{Pozzorini:2020hkx,Lang:2020nnl},
renormalised amplitudes in $D$ dimensions can be obtained from amplitudes
in $\Dn=4$ dimensions through a general formula of the form
\bea 
\mathbf{R}\barM_{2}(\heli) &=& 
\calM_{2}(\heli) + 
\calMCT_{1,1}(\heli) +
\calMCT_{0,2}(\heli) +
\calMCT_{0,1,1}(\heli){}\,.
\label{eq:masterformula2}
\eea
Here the first term on the rhs is the unrenormalised two-loop
amplitude in $\Dn=4$,
while each of the three additional contributions 
embodies standard counterterms for the subtraction of 
UV divergences in combination with rational counterterms for the 
reconstruction of the contributions of the $(D-4)$-dimensional 
parts of loop numerators. The term
$\calMCT_{1,1}(\heli)$ denotes the one-loop amplitude with all relevant
one-loop counterterm insertions, while 
$\calMCT_{0,2}(\heli)$ and $\calMCT_{0,1,1}(\heli){}$
correspond, respectively, to the tree-level amplitudes with 
single two-loop and double one-loop counterterm insertions.
Similarly as for the related UV counterterms, 
also two-loop rational counterterms of UV origin are 
process-independent~\cite{Pozzorini:2020hkx}.
The explicit expressions for all two-loop rational
counterterms of UV origin in QED and for QCD corrections to 
the full SM have been derived
in~\cite{Pozzorini:2020hkx,Lang:2020nnl,Lang:2021hnw}. 

Two-loop rational terms originating from the interplay of $(D-4)$-dimensional 
numerator parts and IR divergences are currently under investigation.
As anticipated in the introduction, 
we expect that -- at the level of IR-subtracted two-loop amplitudes -- all 
rational terms of IR origin can be cancelled 
by means of a simple and process-independent 
modification of the well-established procedures for the subtraction of two-loop IR divergences.

\subsection{Scattering amplitudes and colour factors}
\label{se:diagsandcolour}

In the original \OpenLoops{} algorithm, as well as in the new algorithm, $L$-loop matrix elements $\calM_{L}$ are computed 
as sums of all Feynman diagrams $\Gamma$ of the scattering process, 
\bea
\label{eq:amplitudes_diasum}
\calM_{L}(\heli)
&=&
\sum\limits_{\Gamma} 
\fullamp{L}{\Gamma}(\heli)\,.
\eea
The contribution from each diagram is factorised
into a colour factor $\colfac{L}{\Gamma}$ and a colour-stripped
helicity amplitude $\amp{L}{\Gamma}(\heli)$,
\bea
\label{eq:amplitudes_colfac} 
\fullamp{L}{\Gamma}(\heli)
&=&
\colfac{L}{\Gamma}\,\amp{L}{\Gamma}(\heli)\,.
\eea
Since each quartic gluon vertex gives rise to three independent colour
structures, Feynman diagrams $\Gamma$ that involve $n_q$ quartic gluon vertices are decomposed into
\bea
\label{eq:quarticcolfact}
\fullamp{L}{\Gamma}
&=&
\sum_{j=1}^{3^{n_q}}\colfac{L}{\Gamma_j}\,\amp{L}{\Gamma_j}\,,
\eea
and, in the following, each colour-factorised contribution $\Gamma_j$
is handled as a different Feynman diagram.

The colour structures $\colfac{L}{\Gamma}$  are algebraically reduced 
to a standard colour basis $\{\calC_i\}$ (see \cite{Buccioni:2019sur} for details), 
\bea
\colfac{L}{\Gamma}&=&\sum\limits_i a_{L,i}(\Gamma)\,\calC_i\,,
\label{eq:colfactdec}
\eea
and Born amplitudes are cast in the form
\bea
\calM_{0}(\heli) \,=\,
\sum\limits_{i} \calA_{0}^{(i)}(\heli)\,\calC_i \,.
\label{eq:colourvector}
\eea
Colour-summed interferences 
are built 
by means of the colour-interference matrix
\bea
\calK_{ij} &=&
\sum\limits_{\col}\,\calC_i^\dagger\,\calC_j\,.
\label{eq:colintB}
\eea
For example, the LO probability density  is computed as
\bea
\calW_{00} &=&
\frac{1}{\nhcs}\sum\limits_{\heli}\sum\limits_{i,j}
\left(\calA_0^{(i)}(\heli)\right)^*\,\calK_{ij}\,
\calA_0^{(j)}(\heli)\,.
\label{eq:bornsqcolint}
\eea
The contribution of an $L$-loop diagram $\Gamma$ with $L\geq 1$ to \eqref{eq:M2WvirtBL} is computed as
\bea
\calW_{0L,\Gamma} &=& 
\frac{1}{\nhcs}\sum_{\helig}\sum_{\col} 2\,\re \Big[\calMstar{0}(\heli)\,
\fullamp{L}{\Gamma}(\helig)\Big] 
\nonumber\\ &=& 
\frac{1}{\nhcs}\,\re\left(\sum_{\helig} 
\calU_{0,\Gamma}(\helig)
\,
\amp{L}{\Gamma}(\helig)\right)\,, \label{eq:M2WvirtBL4Gamma}
\eea
where the colour-stripped loop amplitude
$\amp{L}{\Gamma}(\helig)$
is factorised from the colour--Born interference term
\bea
\calU_{0,\Gamma}(\helig)
&=&
2\,\sum_{\col} \calMstar{0}(\helig)\,\colfac{L}{\Gamma}
\,=\,
2\,\sum_j\bigg[
\sum_i \left(\calA_0^{(i)}(\helig)\right)^*\,\calK_{ij}
\bigg]\,
a_{L,j}(\Gamma).
\label{eq:U0Gamma_def}
\eea
This colour treatment is implemented in the public \OpenLoops{} code~\cite{Buccioni:2019sur}
as well as in the new two-loop algorithm presented in this paper.
%

\subsection{Helicity bookkeeping} \label{sec:helbookkeeping}

In this section we define the helicity labels used in \OpenLoops{} at all loop orders, following the notation of \cite{Buccioni:2017yxi}. 
For the bookkeeping of external momenta and helicities
in a process with $\npart$ scattering particles we introduce the set of particle indices
\bea
\calE=\{1,2,\dots, \npart\}.
\label{eq:fullpartset}
\eea
To characterise the helicity configurations of individual 
particles $p\in\calE$ we use the labels
\bea
\lambda_p=\begin{cases}
\;1,3   & \;\mbox{for fermions with helicity}\;s= -1/2, 1/2\\[2mm] 
\;1,2,3 & \;\mbox{for gauge bosons with}\,s=-1,0,1\\[2mm]
\;0     & \;\mbox{for scalars with}\,s=0 
\end{cases} \qquad\forall\; p\in\calE{}.
\label{eq:hellabellam}
\eea
The configuration 
$\lambda_p=0$ is also used to characterise unpolarised
particles, \ie fermions or gauge bosons whose helicity is still
unassigned at a certain stage of the calculation or has already been summed over. We use a helicity numbering scheme based on the labels
\be
\helibar_{p}=\lambda_p\,4^{p-1},
\label{eq:quatrep}
\ee
which correspond to a quaternary number with $\lambda_p\in \{0,1,2,3\}$ as
$p^{\mathrm{th}}$-last digit and all other digits equal to zero.
This scheme allows us to derive the helicity labels of any set of particles as the sum of the helicity labels of its disjoint subsets, and hence as the sum of the helicity labels
of all its particles. A set of particles $\calE_{a}=\{p_{a_1},\ldots,p_{a_n}\}$ has the helicity label,
\be
\heli_{a}=\sum\limits_{p\in\calE_{a}}\helibar_{p}.
\ee
In particular, the global helicity label of the scattering process is
\be
\helig=\sum\limits_{p\in\calE}\helibar_{p}.
\ee
For two disjoint sets of particles $\calE_{b},\calE_{c}$ the helicity label of the combined set $\calE_a=\calE_{b}\cup\calE_{c}$ is
\be
\heli_a=\heli_b+\heli_c{}.
\ee

\section{Tree and one-loop amplitudes in \OpenLoops{}} \label{sec:treeandoneloopamp}

In this section we review the algorithm for the construction of tree and one-loop amplitudes
that is implemented in the public \OpenLoops{} program. This sets the stage for the discussion of the new two-loop algorithm.

\subsection{Tree-level amplitudes} \label{sec:treeamp}

At tree level, the colour-stripped amplitude of a Feynman diagram $\Gamma$ is decomposed into
two subtrees connected by a certain off-shell propagator,\footnote{The Feynman diagrams in this paper are drawn with {\sc Axodraw}~\cite{Vermaseren:1994je}.}
\bea
\amp{0}{\Gamma}(\heli)
&=&
\vcenter{\hbox{\scalebox{1.}{\includegraphics[width=35mm]{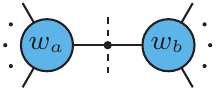}}}}
\,=\,
w^{\sigma_a}_a(k_a,h_a)
\,\delta_{\sigma_a\sigma_b}
\widetilde{w}^{\sigma_b}_b(k_b,h_b)
\,.
\label{eq:cuttreeA}
\eea
Here $\momk{a}=-\momk{b}$ and $\sigma_a,\sigma_b$ 
are the off-shell momenta and Lorentz/spinor indices\footnote{Spinor and Lorentz indices have values from $1$ to $4$
while the indices associated with a scalar propagator only have the value $1$.}
of the subtrees, while
$h_a,h_b$ denote the helicity labels of the subsets of external on-shell 
particles connected to the respective subtrees, and
\be
\helig = \heli_a + \heli_b
\ee
corresponds to the global helicity of all scattering particles.
The subtree $w_a$ is bounded by external wave functions and the off-shell propagator,
at which the diagram was cut,
while $\widetilde{w}_b$ is 
either a single external wave function or is
bounded by a set of external wave functions and a vertex that connects 
it to $w_a$.
The relevant subtrees are 
generated in a recursive way
starting from the external wave functions, and 
connecting them to other external subtrees
through operations of the form
\bea
w^{\sigma_a}_a(k_a,h_a)
\;&=\;\;
\vcenter{\hbox{\scalebox{.8}{\includegraphics[width=60mm]{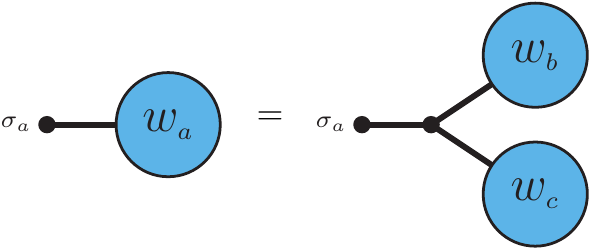}}}}
\;\\&=\,
\displaystyle
\frac{X_{\sigma_b\sigma_c}^{\sigma_a}(\momk{b},\momk{c})}{\momk{a}^2-\mass{a}^2}\;
w^{\sigma_b}_b(\momk{b},\heli_{b})\;\nonumber
w^{\sigma_c}_c(\momk{c},\heli_{c}){}\,,
\label{eq:treerecursionA}
\eea
leading to subtrees with an increasing number of external particles.
The tensor $X_{\sigma_b\sigma_c}^{\sigma_a}$ corresponds to the 
triple vertex that connects  $w_b, w_c$ together with
the numerator of the off-shell propagator 
with momentum $k_a=k_b+k_c$. Each step is performed for all independent helicity configurations $\heli_{b}, \heli_{c}$, and $\heli_a = \heli_b +\heli_c$.
For quartic vertices the relation
\be
\label{eq:treerecursionB}
w^{\sigma_a}_a(\momk{a},\heli_{a}) =
\f{X_{\sigma_b\sigma_c\sigma_d}^{\sigma_a}(\momk{b},\momk{c},\momk{d})}{\momk{a}^2-\mass{a}^2}\; 
w^{\sigma_b}_b(\momk{b},\heli_{b})\;
w^{\sigma_c}_c(\momk{c},\heli_{c}){}\;
w^{\sigma_d}_d(\momk{d},\heli_{d})
\ee
with $\heli_a = \heli_b +\heli_c+\heli_d$ and $k_a=k_b+k_c+k_d$ is used.
The tree recursion is implemented in such a way that
the subtrees contributing to multiple Feynman diagrams are computed only once.

\subsection{One-loop amplitudes} \label{sec:loopamp}

The colour-stripped amplitude of a one-loop diagram $\Gamma$ is given by\footnote{We suppress the argument $\Gamma$ in $\calN$ to simplify the notation.}
\bea
\amp{1}{\Gamma}(\heli)\;&=&\;
\int\!\rd\momq_1\,\f{\Tr\Big[{\calN^{(1)}}(q_1,\heli)\Big]}{\Dbi{0} \Dbi{1}\cdots \Dbi{N_1-1}}
\,=\,
\vcenter{\hbox{\scalebox{.9}{\includegraphics[height=33mm]{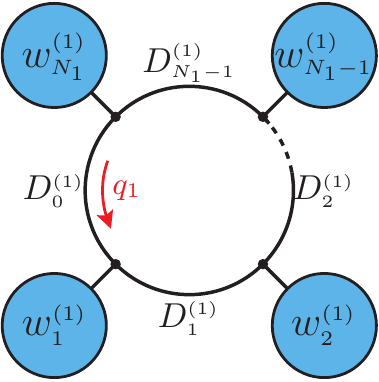}}}}\quad{}.
\label{eq:OLfulldia}
\eea
Here the loop numerator ${\calN^{(1)}}(q_1,\heli)$ and its 
fundamental building blocks, \ie the loop momentum $q_i$, Dirac matrices 
${\gamma}^{\mu}$ and metric tensors ${g}^{\mu\nu}$,
are all evaluated in four dimensions. 
For such four-dimensional quantities and amplitudes computed in $\Dn=4$ dimensions we consistently use 
symbols without a bar.\footnote{See Section~2.1 
of~\cite{Pozzorini:2020hkx} for more details on our
conventions for quantities in $\Dn=D$ and $\Dn=4$ dimensions.
} 
The trace in \eqref{eq:OLfulldia} represents the contraction of Lorentz/spinor 
indices along the loop.
The scalar denominators 
\be
\Dbi{a}(\momq_1)=(\momq_1 +
\momp{1a})^2-\mass{1a}^2
\ee
are kept in $D$ dimensions. They depend
on the $D$-dimensional loop momentum $\momq_1$, the internal mass
$m_{1a}$, and
\be
p_{1a} = \sum\limits_{b=1}^{a} k_{1b}\,,
\ee
where $k_{1b}$ is the external momentum entering the loop between $\Dbi{b-1}$ and $\Dbi{b}$.

For later convenience we apply a label $(i)$ to various $q_i$-dependent building blocks of a loop integrand, where $q_i$ is either a loop momentum or a linear combination of independent loop momenta. At one-loop level, $i=1$, since there is only a single loop momentum.
For the integration measure in loop-momentum space we define the shorthand
\bea
\int\!\rd\momq_i & = & \mu^{2\eps} \int \f{\rd^{^D}\! \bar
q_i}{(2\pi)^{^D}}\,,
\eea
where $\mu$ is the scale of dimensional regularisation.

In \OpenLoops{} the loop numerator is constructed through a numerical recursion that exploits its factorisation into loop segments. A loop segment consists of a loop propagator, one adjacent triple or quartic vertex, and the one or two external subtrees connected to this loop vertex.
In the case of a triple vertex the loop segment has the form
\bea
\Big[\segment{i}{a}(q_i,\helisegment{i}{a})\Big]_{\indc{i}{a-1}}^{\indc{i}{a}}
&=& \!\!
\raisebox{3mm}{\parbox{22mm}{
\scalebox{.9}{\includegraphics[height=19mm]{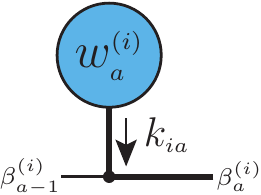}}}}= 
\,\Bigg\{\Big[Y^{\sigma}_{ia}\Big]_{\indc{i}{a-1}}^{\indc{i}{a}}\!\! 
+ \Big[Z_{ia,\nu}^{\sigma}\Big]_{\indc{i}{a-1}}^{\indc{i}{a}}\!\!\; q_i^\nu 
\Bigg\}\, w^{(i)}_{a \sigma}(\momk{ia},\helisegment{i}{a})\,,
\label{eq:seg3point}
\eea
where an external subtree $w_a^{(i)}$ is connected to a loop vertex and 
the adjacent loop propagator. The index $a\in[1,N_i]$ corresponds to the position of the segment along the loop.
In renormalisable theories, a segment
can be written as a rank-one polynomial in the loop momentum
with coefficients $Y$ and $Z$. The indices $\indc{i}{a}$ are, depending on the particle type in the loop propagator, 
Lorentz indices (gauge bosons) or spinor indices (fermions) with $\indc{i}{a}=1,\ldots,4$. For scalar particles (ghosts, scalars) the index has a fixed value $\indc{i}{a}=1$.
Segments associated with a quartic vertex involve two subtrees, $w_{a_1}^{(i)}$ and $w_{a_2}^{(i)}$, and are of rank zero in $q_1$,
\be
\Big[\segment{i}{a}(q_i,\helisegment{i}{a})\Big]_{\indc{i}{a-1}}^{\indc{i}{a}}
=\quad
\raisebox{3mm}{\parbox{29mm}{
\includegraphics[height=17mm]{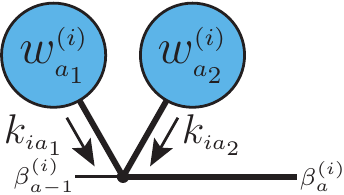}}}
=\, \Big[Y^{\sigma_1\sigma_2}_{ia}\Big]_{\indc{i}{a-1}}^{\indc{i}{a}}
\, w^{(i)}_{a_1\sigma_{1}}(\momk{ia_1},\helisegment{i}{a_1})
\, w^{(i)}_{a_2\sigma_{2}}(\momk{ia_2},\helisegment{i}{a_2})
\label{eq:seg4point}
\ee 
with $\helisegment{i}{a}=\helisegment{i}{a_1}+\helisegment{i}{a_2}$ and  $k_{ia}=k_{ia_1}+k_{ia_2}$.

At one loop, the numerator is computed by cut-opening the loop
at one propagator, which results in a tree-like object
consisting of a product of $N_1$ loop segments,
\bea
\Bigg[\calN^{(1)}(q_1,\heli)\Bigg]_{\indc{1}{0}}^{\indc{1}{N_1}} &=&
\vcenter{\hbox{\scalebox{.9}{\includegraphics[height=33mm]{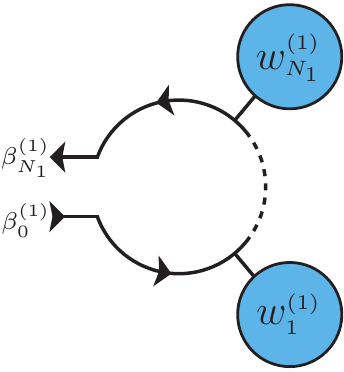}}}}\label{eq:oneloopfac}
\\ &=&
\Bigg[\segment{1}{1}(q_1,\helisegment{1}{1})\Bigg]_{\indc{1}{0}}^{\indc{1}{1}}\,
\Bigg[\segment{1}{2}(q_1,\helisegment{1}{2})\Bigg]_{\indc{1}{1}}^{\indc{1}{2}}\cdots\;
\Bigg[\segment{1}{N_1}(q_1,\helisegment{1}{N_1})\Bigg]_{\indc{1}{N_1-1}}^{\indc{1}{N_1}}, 
\nonumber
\eea
where $\indc{1}{0}, \indc{1}{N_1}$ are the Lorentz/spinor indices of the cut propagator, and $\helisegment{1}{a}$ describes the  helicity configuration of the $a$-th subtree,
\be
\helisegment{1}{a} = \sum\limits_{p \in \particlesetsegment{i}{a}} \helibar_{p}{}, \label{eq:heli_segment}
\ee
where $\particlesetsegment{i}{a}\subseteq\calE{}$ is the 
corresponding set of external particles.

The loop numerator is constructed through recursive matrix multiplications, 
\bea
\calN_n^{(1)}(q_1,\helipc{1}{n}) &=& \calN^{(1)}_{n-1}(q_1,\helipc{1}{n-1})\segment{1}{n}(q_1,\helisegment{1}{n})\,, 
\label{eq:OLrec}
\eea
which are applied for $n=1,\dots, N_1$, starting from
the initial condition $\calN_{0}=\idop$.
The label 
\be
\helipc{1}{n} = \sum\limits_{a=1}^{n} \helisegment{1}{a} \label{eq:heli_partial_chain}
\ee
describes the helicity configuration of the external legs entering the first $n$
segments, and $\helipc{1}{n} = \helipc{1}{n-1} + \helisegment{1}{n}$.
The operations \eqref{eq:OLrec} are referred to as dressing steps, and the partially dressed numerator
\bea 
\calN_{n}^{(1)}(q_1,\helipc{1}{n}) &=&
\prod\limits_{a=1}^{n}\segment{1}{a}(q_1,\helisegment{1}{a})
\, = \,
\raisebox{3mm}{\parbox{75mm}{
\scalebox{.9}{\includegraphics[height=19mm]{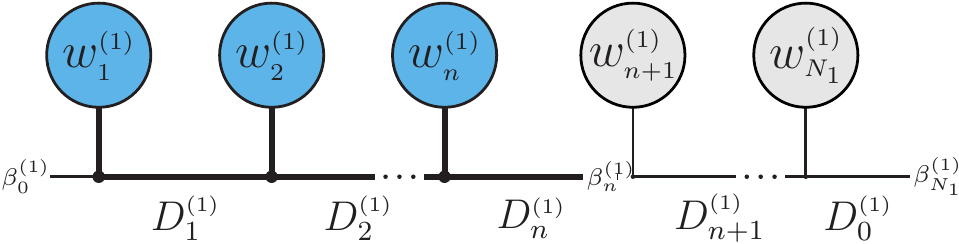}}}}
\eea 
is called an open loop.
In this schematic representation, blue and grey blobs correspond
to the dressed loop segments
and those that remain to be dressed, respectively.
Each open loop is a polynomial in $q_1$,
\bea \calN_{n}^{(1)}(q_1,\helipc{1}{n}) &=&
\sum\limits_{r=0}^R
\calN^{(1)}_{n;\,\mu_1\dots\mu_r}(\helipc{1}{n})
\,q_1^{\mu_1}\cdots q_1^{\mu_r}\,, 
\label{eq:partnum}
\eea 
with rank $R \leq n$, and the dressing recursion \eqref{eq:OLrec} is implemented at the 
level of the tensor coefficients $\calN^{(1)}_{n;\,\mu_1\dots\mu_r}$. The explicit form of a dressing step for a segment \eqref{eq:seg3point} with a three-point vertex is
\bea
\Big[\calN^{(1)}_{n;\,\mu_1\dots\mu_r}(\helipc{1}{n})\Big]_{\indc{1}{0}}^{\indc{1}{n}}
&=&
\Bigg\{
\Big[\calN^{(1)}_{n-1;\,\mu_1\dots\mu_r}(\helipc{1}{n-1})\Big]_{\indc{1}{0}}^{\indc{1}{n-1}}
\Big[Y_{1n}^{\sigma}\Big]_{\indc{1}{n-1}}^{\indc{1}{n}}
\nonumber\\
&&+
\Big[\calN^{(1)}_{n-1;\,\mu_2\dots\mu_r}(\helipc{1}{n-1})\Big]_{\indc{1}{0}}^{\indc{1}{n-1}}
\Big[Z_{1n,\mu_1}^{\sigma}\Big]_{\indc{1}{n-1}}^{\indc{1}{n}}
\Bigg\}
\,w_{n\sigma}^{(1)}(\momk{n},\helisegment{1}{n}).\qquad
\label{eq:OLrecfullcoeffa}
\eea

For an efficient implementation the $\mu_1\dots\mu_r$ indices are symmetrised. 
In the final step of the dressing algorithm, the trace is taken over the indices $\indc{1}{0}, \indc{1}{N_1}$,
\be
\calN^{(1)}_{\mu_1\dots\mu_r}(\helig)=\Tr\Big[\calN^{(1)}_{N_1;\,\mu_1\dots\mu_r}(\helig)\Big],
\label{eq:onelooptrace}
\ee
and for the amplitude of the colour-stripped Feynman diagram 
\refeq{eq:OLfulldia}
we obtain
\bea
\amp{1}{\Gamma}(\heli)\;&=&\;
\sum\limits_{r=0}^N
\calN^{(1)}_{\mu_1\dots\mu_r}(\helig)
\int\!\rd\momq_1\,
\f{q_1^{\mu_1}\ldots q_1^{\mu_{r}}}{ 
D_{0}\!\cdots\! D_{N-1}}.\label{eq:oneloopTIcont}
\eea
The tensor integrals on the rhs can be reduced with external 
libraries such as \Collier~\cite{Denner:2016kdg}.
Alternatively, they can be reduced with the on-the-fly method of \cite{Buccioni:2017yxi},
where dressing steps are interleaved with reduction steps in such a way that the tensor rank remains low at all stages of the calculation. The on-the-fly reduction is the default method 
in \OpenLoopstwo{} \cite{Buccioni:2019sur}.

\subsection{Born-loop interference} \label{sec:loopampU}

As pointed out in \cite{Buccioni:2017yxi}, for the efficient construction of the helicity- and colour-summed Born-loop interference defined in \eqref{eq:M2WvirtBL4Gamma} 
it is convenient to absorb the colour--Born interference factor \eqref{eq:U0Gamma_def} 
into the loop numerator.
In this approach, instead of the original helicity-dependent numerator we construct the helicity summed quantity
\be
\calU(q_1) = \sum_{\helig} \calU_0(\helig)\,\calN^{(1)}(q_1,\helig) 
= \sum_{\helig} \calU_0(\helig) \prod_{a=1}^{N_1} \segment{1}{a}(\helisegment{1}{a})
\label{eq:colborninterf1l}
\ee
with the colour--Born interference 
\be
\calU_0(\helig)=2 \sum_{\col}
\calMstar{0}(\helig)
\,\colfac{1}{\Gamma}\,,
\label{eq:OLrecUinit}
\ee
where, again, the label $\Gamma$ is kept implicit.
Exploiting the factorisation of $\calU(q_1)$ into $\calU_0(h)$ and loop segments, one can write
\be
\calU(q_1) = 
\sum\limits_{\helisegment{1}{N_1}} 
\Bigg[
\;\ldots\;
\sum\limits_{\helisegment{1}{2}}
\Bigg[\sum\limits_{\helisegment{1}{1}}\calU_0(\helig)\,\segment{1}{1}(\helisegment{1}{1})
\Bigg]\segment{1}{2}(\helisegment{1}{2})
\;\ldots\;
\Bigg]
\segment{1}{N_1}(\helisegment{1}{N_1}),
\ee
where helicity sums are partially factorised at the level of individual segments.
More precisely, each time the numerator is dressed with a new segment
$\segment{1}{n}(q_1,\helisegment{1}{n})$, the helicity d.o.f.~of the associated subtree
$w^{(1)}_n(\helisegment{1}{n})$ 
can be summed, and the subsequent dressing steps depend only on the helicities of the yet undressed segments. 
This way of constructing the loop numerator \eqref{eq:colborninterf1l} corresponds to a recursive dressing algorithm
\bea
\calU_n(q_1,\helipcc{1}{n}) &=& \sum\limits_{\helisegment{1}{n}}
\calU_{n-1}(q_1,\helipcc{1}{n-1})\,
\segment{1}{n}(q_1,\helisegment{1}{n})\,,
\label{eq:OLrecU}
\eea
where the initial condition \eqref{eq:OLrecUinit} is used, and the helicity label
\be
\helipcc{1}{n}\,=\, \helig - \helipc{1}{n} = \sum\limits_{a=n+1}^{N_1}\helisegment{1}{a}
\label{eq:heldef_pdcc_g_1l}
\ee
corresponds to the helicity configuration of all
undressed segments. 
To keep track of the polynomial 
dependence of the partially dressed numerators
on the loop momentum $q_1$,
a tensorial representation similar to \eqref{eq:partnum} is used,
\bea \calU_{n}(q_1,\helipcc{1}{n}) &=&
\sum\limits_{r=0}^R
\calU_{n;\,\mu_1\dots\mu_r}(\helipcc{1}{n})
\,q_1^{\mu_1}\cdots q_1^{\mu_r}\,, 
\label{eq:partnumU}
\eea 
and the dressing recursion is implemented at the level of 
tensor coefficients in the same form as 
in \eqref{eq:OLrecfullcoeffa}, but 
with an additional sum over $\helisegment{1}{n}$
as in~\refeq{eq:OLrecU}.
This on-the-fly helicity summation approach
guarantees the smallest possible number of helicity configurations in the last dressing steps, where the rank in
$q_1$,
and hence the number of tensor coefficients 
and required matrix multiplications,
is highest.

At the end of the dressing recursion, \ie after step $n=N_1$,
in analogy with~\eqref{eq:onelooptrace} we take the trace
\be
\calU_{\mu_1\dots\mu_r}=\Tr\Big[\calU_{N_1;\,\mu_1\dots\mu_r}(\helipcc{1}{N_1})\Big],
\label{eq:onelooptraceU}
\ee
where $\helipcc{1}{N_1}=0$ since no undressed segments are left.
Finally,
combining the tensor coefficients with the associated tensor integrals,
one arrives at
\bea
\calW_{01,\Gamma} &=& \frac{1}{\nhcs}\,\mathrm{Re}\left[
\sum\limits_{r=0}^N
\calU_{\mu_1\dots\mu_r}
\int\!\rd\momq_1\,
\f{q_1^{\mu_1}\ldots q_1^{\mu_{r}}}{ 
D_{0}\!\cdots\! D_{N-1}}
\right]\,,
\label{eq:ofrscattdens}
\eea
which corresponds to the one-loop virtual scattering probability
density as defined in~\eqref{eq:M2WvirtBL4Gamma}.

The reduction of the tensor integrals can be performed with external 
libraries, such as \Collier~\cite{Denner:2016kdg}, or with the on-the-fly 
reduction method of \cite{Buccioni:2017yxi}.


\begin{figure}[t]
 \begin{tabular}{ccccc}
  $\vcenter{\hbox{\scalebox{1.}{\includegraphics[width=0.3\textwidth]{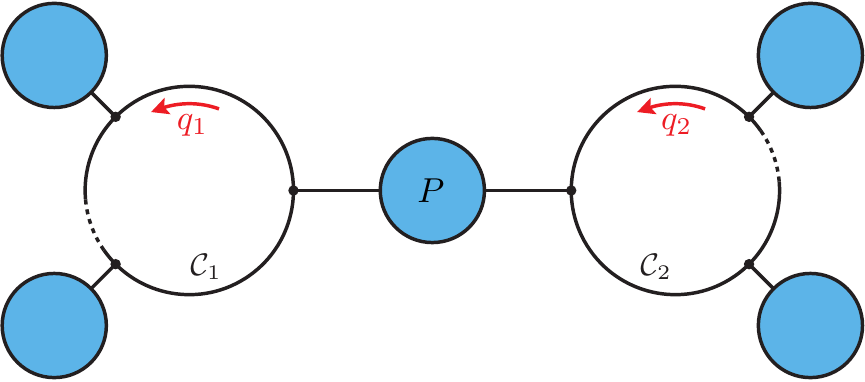}}}}$ 
& \quad\qquad &
  $\vcenter{\hbox{\scalebox{1.}{\includegraphics[width=0.2\textwidth]{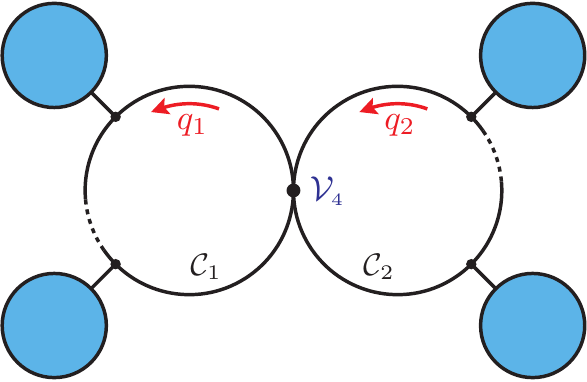}}}}$ 
& \quad\qquad &
  $\vcenter{\hbox{\scalebox{1.}{\includegraphics[width=0.3\textwidth]{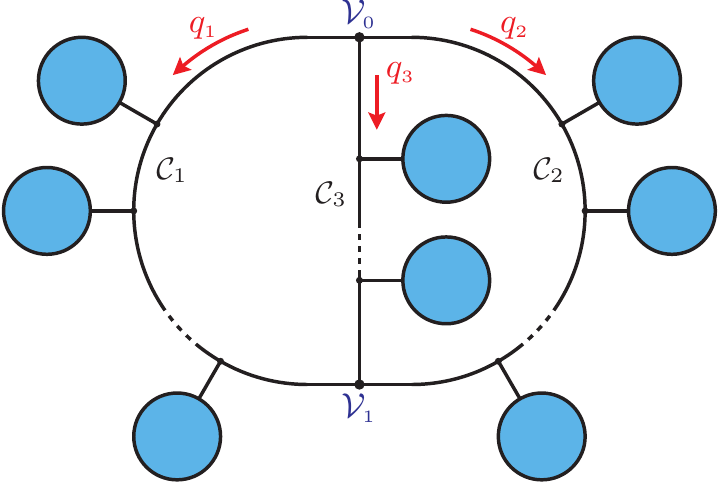}}}}$ 
\\
(\redtwo) && (\redone) & &  (Irred)
 \end{tabular}
\caption{Categorisation of two-loop topologies into reducible (Red)
and irreducible (Irred) ones. Reducible two-loop topologies are further split
into two subcategories, where two one-loop subdiagrams 
are either connected to a bridge through two vertices (\redtwo) 
or are attached to each other through a common quartic vertex (\redone).
Similarly as in~\refse{sec:treeandoneloopamp} the blue blobs 
denote tree structures that are connected to internal and 
external lines (the latter are not shown). 
In the~\redtwo~topology the blue blob labelled $P$ 
can be connected only 
to the two visible internal lines or also to 
additional external lines.
\label{fig:twoloopclasses}}
\end{figure}

\section{Reducible two-loop integrands} 
\label{sec:twoloopred}

In this and the following section we present a new automated algorithm that
extends the construction of the tensor coefficients~\refeq{eq:onelooptraceU}
to two loops. Specifically, we will focus on the construction of tensor
coefficients for unrenormalised two-loop amplitudes in $\Dn=4$ dimensions, \ie~with four-dimensional
numerators, while all other ingredients of renormalised two-loop
amplitudes~\eqref{eq:masterformula2} can be obtained from one-loop and tree
amplitudes with UV and rational counterterm insertions.
All amplitudes are split into individual Feynman diagrams, 
and their colour structures are factorised as described in
\refeq{eq:amplitudes_diasum}--\refeq{eq:quarticcolfact}.
Thus we will concentrate on the construction of tensor coefficients
for colour-stripped  unrenormalised two-loop diagrams.

At two loops we categorise Feynman diagrams into reducible and irreducible ones
as illustrated in \reffi{fig:twoloopclasses}. This 
categorisation is based on the structure of 
loop chains. Such chains consist of loop propagators that are linked to each other
and to external subtrees through connecting vertices. 
More precisely, a loop chain $\calC_i$ contains the propagators that 
depend only on a certain loop momentum $q_i$ when
all external momenta are set to zero.

Two-loop diagrams that involve only two loop chains $\calC_1, \calC_2$,
with two independent loop momenta $q_1, q_2$, are categorised as
reducible.  This category is further split into two-loop diagrams of type
\redtwo~and \redone~depending on how the two chains are connected to each
other.  As indicated in \reffi{fig:twoloopclasses}, in reducible two-loop
diagrams of type \redtwo~the chains $\calC_1$ and $\calC_2$ are connected
through a tree structure $P$, which is referred to as bridge, while the
type \redone~corresponds to the case where the two chains are connected
through a single quartic vertex $\vertex{4}$.
Since reducible two-loop diagrams factorise into two
one-loop subdiagrams, for their calculation 
one can exploit various functions of the one-loop
\OpenLoops~framework.
An algorithm for the construction of 
reducible two-loop integrands of type \redtwo~and~\redone~is
presented in \refses{sec:twoloopred2}{sec:twoloopred1}.

Two-loop diagrams that involve three loop chains $\calC_1, \calC_2, \calC_3$
are categorised as irreducible.  In this case the three chains are connected
to each other by two vertices $\vertex{0}, \vertex{1}$, and of the three loop
momenta $q_1, q_2, q_3$ only two are linearly independent. 
The loop momenta can always be chosen 
in such a way that $q_1+q_2+q_3=0$. 
An algorithm for the construction of irreducible two-loop integrands is
presented in~\refse{sec:twoloopirred}.

In general, the algorithms presented in
\refses{sec:twoloopred2}{sec:twoloopred1} and~\ref{sec:twoloopirred}
compute tensor coefficients that corresponds to the
helicity and colour-summed interference \eqref{eq:M2WvirtBL4Gamma}
of the full Born amplitude with the
integrand of individual two-loop diagrams.

\subsection{Reducible two-loop integrands of class \redtwo} \label{sec:twoloopred2}

A reducible two-loop diagram of class \redtwo~factorises into two chains $\calCh{i}$ and a bridge $P$ as depicted in
\reffi{fig:twoloopclasses}.
The loop chains $\calCh{i}$ each consist of $N_i$ loop propagators ($i=1,2$), and
the external subtrees $\subtree{i}{1},\ldots,\subtree{i}{N_i-1}$ connected to the triple or quartic loop vertices as in \eqref{eq:seg3point} and \eqref{eq:seg4point}. The bridge $P$ is a tree structure connected to each of the two chains with a corresponding vertex, which we 
call the bridge vertex of that chain. In general, the bridge is also connected to a subset of the external particles of the 
scattering process.
The colour-stripped amplitude of a reducible diagram $\Gamma$
has the generic form
\bea \mathcal{A}_{2,\Gamma}(\helig)\;&=&\quad
\vcenter{\hbox{\scalebox{.9}{\includegraphics[width=0.65\textwidth]{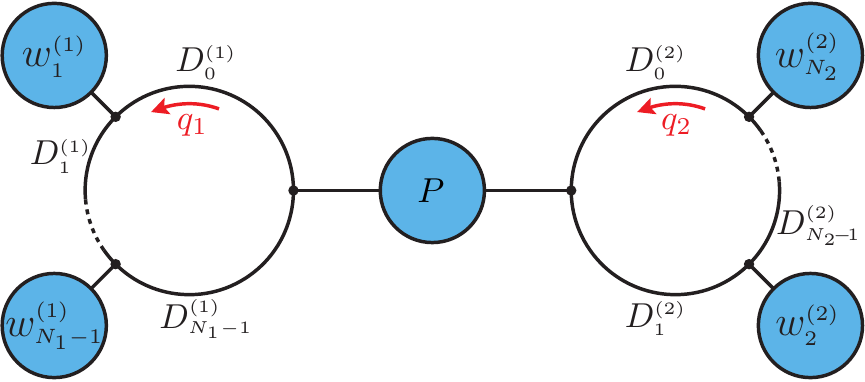}}}}
\nonumber\\[4mm]
&=&\quad\int\!\rd\momq_1\!
\f{\Tr\Big[{\calN^{(1)}}(q_1,\helic{1})\Big]^{\alpha_1}   }{\denc{1}} 
\,P_{\alpha_1\alpha_2}(\helic{\rB})\,
\int\!\rd\momq_2
  \f{ \Tr\Big[{\calN^{(2)}}(q_2,\helic{2})\Big]^{\alpha_2}}{\denc{2}}\,,    
\label{eq:A2dred}
\eea
with the denominator chains
\bea
\label{eq:dendef2l_red}
\denc{i}&=&
D^{(i)}_0(\bar q_i)\cdots
D^{(i)}_{N_i-1}(\bar q_i)\,,
\qquad
D^{(i)}_a(\bar q_i) \,=\, \left(\bar q_i + p_{ia}\right)^2-m_{ia}^2\,,
\eea
where $p_{ia}$ and $m_{ia}$ are the external momenta and internal masses along the chain $\calC_{i}$. 
The Lorentz/spinor indices  $\alpha_{1},\alpha_2$ connect the bridge
$P_{\alpha_1\alpha_2}$ to the chains $\calC_1,\calC_2$.
The external subtrees $w_a^{(i)}$ depicted in \eqref{eq:A2dred} play the
two-fold role of single subtrees or pairs of subtrees connected to the loop
chains via triple and quartic vertices, respectively.
The global helicity configuration is given by
\be
\helig = \helic{1} + \helic{\rB} + \helic{2},
\ee
with the chain helicities 
\be
\helic{i}=\sum\limits_{a=1}^{N_i} \helisegment{i}{a}\quad (i=1,2)\,,
\ee
and the bridge helicity $\helic{\rB}$,
which corresponds to the  helicity configuration of all
external particles connected to $P$. 

Reducible two-loop diagrams can be efficiently constructed by the following algorithm, 
which uses and extends functions of the one-loop \OpenLoops{} algorithm.

\paragraph{Ordering and cutting rule:}
 \begin{figure}[t]
\begin{center}
\includegraphics[width=0.65\textwidth]{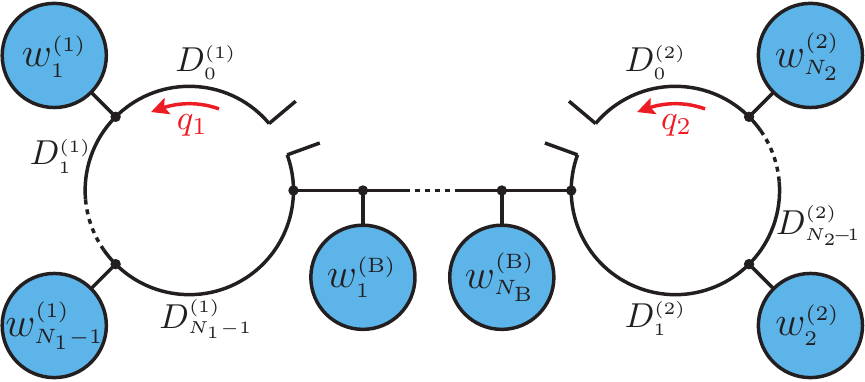} 
 \end{center}
 \caption{Cut open reducible two-loop diagram of type \redtwo.}
 \label{fig:openRD}
\end{figure}
 Chain $\calCh{1}$ is chosen to be the chain with less loop
propagators.\footnote{%
In case there is more than one chain 
with the minimum number of propagators, then the one with 
the minimum number of helicity d.o.f.~is chosen to be 
$\calC_1$. If in turn there is more then one chain with the minimum number
of propagators and helicity d.o.f., then $\calC_1$ is selected based on the
particle content of the external subtrees that are connected to the various chains.
}
As illustrated in \reffi{fig:openRD}, the construction of the loop numerator is organised by cut-opening the two one-loop subdiagrams between the bridge vertices and the $D_0^{(i)}$
propagators.
As usual the corresponding external momenta are set to $p_{i0}=0$.
The position of the cut (left or right of the bridge vertex)
is determined in a unique way based on the properties of the subtrees. The dressing directions of the loop chains 
correspond to the direction of the loop momenta $q_i$.

\paragraph{Step~1 -- Construction of chain $\calCh{1}$:} 
The numerator of the chain $\calCh{1}$ factorises into loop segments,
\be
\Tr\Big[\numc{1}(q_1,\helic{1})\Big]^{\alpha_1}= 
\segment{1}{0}(q_1)\,\,
\segment{1}{1}(q_1,\helisegment{1}{1})\cdots
\segment{1}{N_1-1}(q_1,\helisegment{1}{N_1-1})
\Big[\segment{1}{N_1}(q_1,\helisegment{1}{N_1})\Big]_{}^{\alpha_1},
\label{eq:numcred1}
\ee
where the last segment carries the Lorentz/spinor index $\alpha_{1}$ that connects it to the bridge, while the indices connecting the segments to each other are kept implicit.

The recursive construction of \eqref{eq:numcred1} starts from the segment $\segment{1}{0}$, which is simply the propagator numerator corresponding to $D_0^{(1)}$.
The segments $\segment{1}{a}$ for \mbox{$a=1,\ldots,N_1-1$} consist of a loop vertex together with the attached subtree(s) $\subtree{1}{a}$, and the attached propagator numerator corresponding to $D_a^{(1)}$. The last segment $\segment{1}{N_1}$ consists of the bridge vertex on chain $\calCh{1}$, and in the case of a four-vertex a subtree $\subtree{1}{N_1}$. This decomposition ensures that the open index $\alpha_{1}$ is only introduced in the last step of the construction of $\calCh{1}$.

The first chain is constructed starting from the initial condition $\numpc{1}{-1}=\idop$ with dressing steps
 \be
 \numpc{1}{n}(q_1,\helipc{1}{n}) = \numpc{1}{n-1}(q_1,\helipc{1}{n-1})\cdot\segment{1}{n}(q_1,\helisegment{1}{n})
\qquad\text{for}\quad n=0,\ldots,N_1\;{},
 \ee
where we use the partial chain helicities defined in \eqref{eq:heli_partial_chain} with 
$\helipc{1}{n} = \helipc{1}{n-1} + \helisegment{1}{n}$.
Similarly as in 
\eqref{eq:OLrec}--\eqref{eq:onelooptrace}
the recursion is implemented in terms of tensor coefficients and, taking the trace after the last step, yields the coefficients
\be
\Big[\calN^{(1)}_{\mu_1\dots\mu_r}(\helic{1})\Big]^{\alpha_1}=\Tr\Big[\calN^{(1)}_{N_1;\,\mu_1\dots\mu_r}(\helipc{1}{N_1})\Big]^{\alpha_1},
\label{eq:onelooptraceCh1red}
\ee
where $\helic{1}=\helipc{1}{N_1}$.
Contracting all coefficients with corresponding one-loop tensor integrals
results into the closed first loop
\bea
\label{eq:closed1stloop}
\Big[P_{-1}(\helic{1})\Big]^{\alpha_1}
&=&
\int\!\rd\momq_1\!
\f{\Tr\Big[{\numc{1}}(q_1,\helic{1})\Big]^{\alpha_1}   }{\denc{1}} \nonumber\\
&=&
\sum\limits_{r=0}^{N_1}
\Big[\calN^{(1)}_{\mu_1\dots\mu_r}(\helic{1})\Big]^{\alpha_1}
\int\!\rd\momq_1\,
\f{q_1^{\mu_1}\ldots q_1^{\mu_{r}}}{\denc{1}}
,
\eea
which serves as the starting point for the bridge construction. Choosing $\calCh{1}$ to be the shorter chain, \ie $N_1 \leq N_2$, ensures that the number of helicity states $\helic{1}$, for which the tensor coefficients need to be computed and contracted with a tensor
integral, is minimal. 
Step~1 is performed for all diagrams,
recycling partially or entirely constructed chains wherever possible.

\paragraph{Step 2 -- Bridge construction:}

The bridge involves various segments $\segment{\rB}{a}$ with
\mbox{$a=0,\ldots,N_{\rB}$} as depicted in \reffi{fig:openRD}.  
As a starting point for its construction,
the segment $\segment{\rB}{0}$, which consists solely of the first bridge propagator, 
is attached to the one-loop subdiagram~\refeq{eq:closed1stloop},
\bea
P_{0}(\helipc{\rB}{0}) &=&
P_{-1}(\helipc{\rB}{0}) \cdot \segment{\rB}{0}. 
\label{eq:brigdeIC}
\eea
Here the global helicity $\helic{1}$ of the subdiagram~\refeq{eq:closed1stloop}
has been renamed as $\helipc{\rB}{0}$.
The remaining bridge segments $\segment{\rB}{n}(\helisegment{\rB}{n})$ 
with $n>0$ consist of a
vertex and a bridge propagator together with one or two external subtrees
$\subtree{\rB}{n}$. Note that, in contrast to loop segments, these tree segments
contain also the associated propagator denominators, 
which depend solely on external momenta and internal masses.
The label $\helisegment{\rB}{n}$ corresponds
to the helicity configuration of the external particles in $\subtree{\rB}{n}$.
These subtrees are recursively attached to~\refeq{eq:brigdeIC}
through dressing steps
\bea
P_{n}(\helipc{\rB}{n}) &=&
P_{n-1}(\helipc{\rB}{n-1}) \cdot \segment{\rB}{n} (\helisegment{\rB}{n})
\label{eq:brigderec}
\eea
for $n=1,\ldots,N_{\rB}$, which 
are implemented in the \OpenLoops{} tree-level algorithm. 
The global helicity configuration for this partially dressed
bridge contracted with the first one-loop subdiagram corresponds to 
\be
\helipc{\rB}{n}\,=\,
\helipc{\rB}{n-1}+\helisegment{\rB}{n}
\,=\,
\helic{1}+\sum\limits_{a=1}^{n}\helisegment{\rB}{a}.
\ee
Recycling opportunities are systematically exploited whenever
partially dressed bridges connected to the first loop occur in multiple diagrams.

The final result of the recursion~\refeq{eq:brigderec} 
corresponds to the first two building blocks on the rhs
of~\refeq{eq:A2dred}, \ie
\bea
\Big[ 
P_{N_\rB}(\helipc{\rB}{})
\Big]_{\alpha_2} 
&=&
\int\!\rd\momq_1\!
\f{\Tr\Big[{\numc{1}}(q_1,\helic{1})\Big]^{\alpha_1}   }{\denc{1}}\,
P_{\alpha_1\alpha_2}(\helic{\rB})\,,
\label{eq:bridgeloopcomb}
\eea
where we define
\be
\helipc{\rB}{}\,=\,
\helipc{\rB}{N_\rB}
\,=\, 
\helic{1} + \helic{\rB}\,.
\ee

\paragraph{Step 3 -- Construction of chain $\calCh{2}$:}

The second chain is constructed starting from the loop segment associated
with the propagator denominator $D_1^{(2)}$. 
In this loop segment we include, as an effective external subtree, the full bridge--loop combination~\refeq{eq:bridgeloopcomb}.
If the bridge and the chain $\calC_2$
are connected by a triple vertex,
according to~\refeq{eq:seg3point}
the first loop segment 
has the form
\bea
\segment{2}{1}(q_2,\helisegment{2}{1})
&=&
\left(Y^{\alpha_2}_{21} + Z^{\alpha_2}_{21,\nu}\,q_2^\nu\right)
\Big[ 
P_{N_\rB}(\helipc{\rB}{})
\Big]_{\alpha_2}\,,
\label{chaintwofstA}
\eea
with $\helisegment{2}{1}=\helipc{\rB}{}$.
Here $Y$ and $Z$ embody the connecting triple vertex 
together with the propagator numerator 
associated with $D_1^{(2)}$.
For a 
quartic vertex, similarly as in~\refeq{eq:seg4point} 
we have
\bea
\segment{2}{1}(q_2,\helisegment{2}{1})
&=&
Y^{\alpha_2\alpha'_2}_{21}
\Big[ 
P_{N_\rB}(\helipc{\rB}{})
\Big]_{\alpha_2}
\Big[w^{(2)}_{1'}(\helisegment{2}{1'})\Big]_{\alpha'_2}
\,,
\label{chaintwofstB}
\eea
where $w^{(2)}_{1'}$ denotes an additional subtree that is 
connected to the quartic vertex, and 
$\helisegment{2}{1}=\helipc{\rB}{}+\helisegment{2}{1'}$.
As a result of~\refeq{chaintwofstA}--\refeq{chaintwofstB} 
the index $\alpha_{2}$ is saturated 
at the beginning of the construction of the 
chain $\calC_2$, which renders the 
subsequent operations more efficient.
The subsequent loop segments $\segment{2}{a}(q_2,\helisegment{2}{a})$ with $a>1$ along $\calC_2$ are defined in the same way as in the one-loop case and have the form \refeq{eq:seg3point} and \refeq{eq:seg4point}.
In this way, the chain of segments $\segment{2}{a}$ corresponds to
\be
\prod\limits_{a=1}^{N_2}\segment{2}{a}(q_2,\helisegment{2}{a})
=
\int\!\rd\momq_1\!
\f{\Tr\Big[{\calN^{(1)}}(q_1,\helic{1})\Big]^{\alpha_1}}{\denc{1}} 
\,P_{\alpha_1\alpha_2}(\helic{\rB})\,
\Big[{\calN^{(2)}}(q_2,\helic{2})\Big]^{\alpha_2}  
\label{eq:numcred2}.
\ee
The dressing steps for this chain are carried out 
at the level of the 
Born--loop interference using the 
one-loop algorithm of \refse{sec:loopampU}.
To this end, using as initial condition 
the colour--Born interference
 \be
\numpii{0}(\helig)
=2\left(\sum_{\col}
\calM^*_0(\helig)
\,\colfac{2}{\Gamma}\right){}\,,
\label{eq:initC2red}
\ee
the various segments are recursively  combined via 
\bea
\numpii{n}(q_2,\helipcc{2}{n})
&=&
\sum\limits_{\helisegment{2}{n}}
\numpii{n-1}(q_2,\helipcc{2}{n-1})\cdot\segment{2}{n}(q_2,\helisegment{2}{n})\,,
\label{eq:dressingstepC2red}
\eea
for $n=1,\ldots,N_2$. In this way helicity states are efficiently summed on-the-fly, and
\be
\helipcc{2}{n} \,=\, \helig - \sum\limits_{a=1}^{n}\helisegment{2}{a}
\,=\,\sum\limits_{a=n+1}^{N_2}\helisegment{2}{a}
\ee
corresponds to the helicity configuration of the segments that are still undressed.
Note that $\helipcc{2}{0} =\helig$\,.
The dressing recursion is again implemented in terms of tensor coefficients,
and after step $n=N_2$ one arrives at the two-loop
scattering probability density along the same lines as
in~\refeq{eq:onelooptraceU}--\refeq{eq:ofrscattdens}, 
\ie
by taking the trace
\be
\calU^{(2)}_{\mu_1\dots\mu_r}=\Tr\Big[\calU^{(2)}_{N_2;\,\mu_1\dots\mu_r}
(\helipcc{2}{N_2})
\Big]\,,
\label{eq:onelooptraceU2}
\ee
where $\helipcc{2}{N_2}=0$, and
combining the tensor coefficients with the associated tensor integrals.
This leads to 
\bea
\calW_{02,\Gamma} &=& \frac{1}{\nhcs}
\,\mathrm{Re}\left[
\sum\limits_{r=0}^{N_2}
\calU^{(2)}_{\mu_1\dots\mu_r}
\int\!\rd\momq_2\,
\f{q_2^{\mu_1}\ldots q_2^{\mu_{r}}}{ 
\denc{2}}\right]\,,
\eea
which corresponds to the two-loop scattering probability
density as defined in~\eqref{eq:M2WvirtBL4Gamma}.

\subsection{Reducible two-loop integrands of class \redone} \label{sec:twoloopred1}
The above algorithm can be easily extended to reducible  diagrams of type
\redone. In this case the one-loop chains $\calC_1, \calC_2$ are connected by a
single four-gluon vertex $\calV_4$ (see \reffi{fig:twoloopclasses}).
The  various colour-stripped
amplitudes that result form the splitting~\refeq{eq:quarticcolfact} of
$\calV_4$ and any other quartic vertices are handled as independent diagrams. For each 
of them the first one-loop subdiagram is constructed 
as in Step~1 of~\refse{sec:twoloopred2}. The only exception is that 
the open index $\alpha_1$, which connects the chain $\calC_1$ to the bridge, is
replaced by two indices $\beta^{(2)}_0$, $\beta^{(2)}_1$.
The latter correspond to the Lorentz indices
that result form cut-opening the two propagators of the chain
$\calC_2$ that are connected to $\calV_4$.
In practice, the quartic vertex $\calV_4$ is included in the 
last segment of the chain $\calC_1$, and the first loop~\refeq{eq:closed1stloop}
assumes the form
\bea
\Big[P_{-1}(\helic{1})\Big]^{\alpha_1}
&\to&
\Big[P_{-1}(\helic{1})\Big]_{\indc{2}{0}}^{\indc{2}{1}}\,.
\eea
Since topologies of type \redone~feature a trivial 
bridge (no bridge segments, $N_{\rB}=\helic{\rB}=0$),
Step~2 can be by-passed. In Step~3 the starting point, \ie 
the first loop segment
of chain $\calC_2$, is simply given by 
\bea
\bigg[\segment{2}{1}(q_2,\helisegment{2}{1})\bigg]_{\indc{2}{0}}^{\indc{2}{1}}
&=&
{}-\ri 
\bigg[P_{-1}(\helic{1})\bigg]_{\indc{2}{0}}^{\indc{2}{1}}\,,
\eea
where $\helisegment{2}{1}=\helic{1}$, and the factor $-\ri$ originates
form the numerator $-\ri g^{\beta\beta'}$ of the 
gluon propagator associated with $D^{(2)}_1$. 
The rest of Step~3 can be implemented 
as in~\refse{sec:twoloopred2}.

\section{Irreducible two-loop integrands} \label{sec:twoloopirred}

The colour-stripped amplitude $\amp{2}{\Gamma}$ of an irreducible two-loop diagram $\Gamma$ 
has the form
\bea \amp{2}{\Gamma} (\heli) \quad&=&\quad
\vcenter{\hbox{\scalebox{1.}{\includegraphics[width=0.5\textwidth]{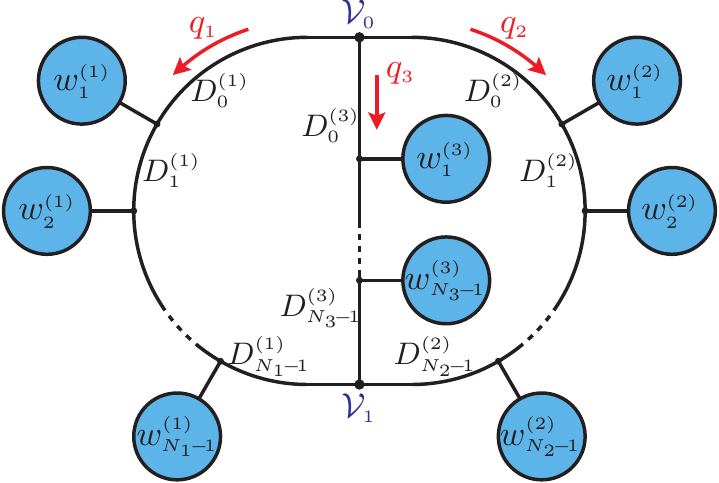}}}}
\nonumber\\[3mm]
&=&\quad\int\!\rd\momq_1\!\int\!\rd\momq_2\,
\f{\calN(q_1,q_2,\heli)
   }{\denc{1}\,\denc{2}\,\denc{3}}\, \bigg|_{\momq_3=-(\momq_1+\momq_2)}\,,   
\label{eq:A2d}
\eea
with the denominator chains
\bea
\label{eq:dendef2l}
\denc{i}&=&
D^{(i)}_0(\bar q_i)\cdots
D^{(i)}_{N_i-1}(\bar q_i)
\qquad\mbox{with}\qquad
D^{(i)}_a(\bar q_i) \,=\, \left(\bar q_i + p_{ia}\right)^2-m_{ia}^2\,,
\eea
where $p_{ia}$ and $m_{ia}$ are the external momenta and internal masses along the chain $\calC_{i}$.
The three chains are connected by two vertices
$\vertex{0},\vertex{1}$, and each connecting vertex $\vertex{a}$ can either be
a three-point vertex, as depicted in~\eqref{eq:A2d}, or a quartic vertex. 
In the latter case $\vertex{a}$ is attached to a an external subtree
$\subtree{V}{a}$, which is not shown in~\eqref{eq:A2d}.
Along each chain $\calC_{i}$ flows a single loop momentum $q_i$ in the
direction from $\vertex{0}$ to $\vertex{1}$, with the boundary condition
$q_3=-(q_1+q_2)$.
If $\vertex{0}$ is a triple
vertex the external momenta associated with the propagators $\Db{i}{0}$ connected 
to $\vertex{0}$ are set to $p_{i0}=0$ for $i=1,2,3$. 
Instead, if $\vertex{0}$ is a quartic vertex 
we choose $p_{10}=p_{20}=0$ and $p_{30}=k_{V0}$,
where $k_{V0}$ is the external
momentum entering through the external subtree $\subtree{V}{0}$ connected to
$\vertex{0}$.

The numerator of~\eqref{eq:A2d} factorizes into three numerator chains 
and the two vertices 
$\vertex{0}, \vertex{1}$, 
each connecting all three chains,
\bea
\calN(q_1,q_2,\heli)&=& \prod\limits_{i=1}^{3}
\Big[\numc{i}(q_i,\helic{i})\Big]_{\indc{i}{0}}^{\indc{i}{N_1}}
   \Big[\vertex{0}(q_1,q_2,q_3,
\helisegment{V}{0})\Big]^{\indci{0}\indcii{0}\indciii{0}} \nonumber\\ && \times
\Big[\vertex{1}(q_1,q_2,q_3,
\helisegment{V}{1})\Big]_{\indci{N_1}\indcii{N_2}\indciii{N_3}}\,
\Bigg|_{q_3=-(q_1+q_2)}\,, \label{eq:numdef2l}
\eea
where the tensors $\vertex{a}(q_1,q_2,q_3,
\helisegment{V}{a})$
represent the connecting vertices $\calV_a$. In  case of quartic vertices
they also embody the related external subtree $w_a^{(V)}(\helisegment{V}{a})$,
thereby inheriting the dependence on the helicity states
$\helisegment{V}{a}$. For triple vertices $\calV_a$ there is no helicity
dependence, and $\helisegment{V}{a}=0$.

The global helicity configuration defined by all external particles of the
two-loop diagram is hence decomposed into
\be
\helig = \sum\limits_{i=1}^{3} \helic{i} + \helisegment{V}{1} +
\helisegment{V}{0}{}\,, \label{eq:heldef_global}
\ee
where $\helic{i}$ denotes the helicity configuration of the chain
$\calC_i$
and is simply given by the sum of the corresponding segment helicities, 
as defined in \eqref{eq:heli_segment},
\be
\helic{i} = \sum\limits_{a=1}^{N_i-1} \helisegment{i}{a}\,.  \label{eq:heldef_chain}
\ee
Each chain numerator in~\eqref{eq:numdef2l} factorises into loop segments
\bea
\numc{i}(q_i,\helic{i}) &=&
\segment{i}{0}(q_i)\,\segment{i}{1}(q_i,\helisegment{i}{1})\,\cdots\,\segment{i}{N_i-1}(q_i,\helisegment{i}{N_i-1})\,. 
\label{eq:numcsimple}
\eea
The segment $\segment{i}{0}(q_i,\helisegment{i}{0})\equiv\segment{i}{0}(q_i)$ consists only of the corresponding propagator
numerator, while all other $\segment{i}{a}(q_i,\helisegment{i}{a})$ with $a\geq 1$ 
are standard loop segments of the form~\refeq{eq:seg3point}
or~\refeq{eq:seg4point}, \ie they consist of a triple or quartic loop vertex
connected to one or two external subtrees and to the
numerator of a loop propagator adjacent to the loop vertex in the 
dressing direction. In loop segments
with a quartic vertex, the external subtrees $\subtree{i}{a}$
depicted in~\eqref{eq:A2d} should be understood as 
pairs of subtrees.

The multiplications in~\eqref{eq:numcsimple} should be understood as matrix
multiplications,
\be
\Big[\numc{i}(q_i,\helic{i})\Big]_{\indc{i}{0}}^{\indc{i}{N_i}} = 
\Big[\segment{i}{0}(q_i)\Big]_{\indc{i}{0}}^{\indc{i}{1}}\,\,
\Big[\segment{i}{1}(q_i,\helisegment{i}{1})\Big]_{\indc{i}{1}}^{\indc{i}{2}}\cdots
\Big[\segment{i}{N_i-1}(q_i,\helisegment{i}{N_i-1})\Big]_{\indc{i}{N_i-1}}^{\indc{i}{N_i}}\,.
\label{eq:numcexpl}
\ee
For simplicity we will suppress the Lorentz/spinor indices $\indc{i}{a}$
wherever possible.

Ultimately, we are interested in the 
construction of the
scattering probability density
\eqref{eq:M2WvirtBL}, 
and in the following we focus on the 
contribution of a single unrenormalised 
two-loop diagram as defined in~\eqref{eq:M2WvirtBL4Gamma}.
To account for the effect of the colour--Born interference,
similarly as in the one-loop relations~\eqref{eq:colborninterf1l} and~\eqref{eq:OLrecUinit}
we introduce 
\be
\calU_0(\helig)\,=\,2\sum_{\col}\calMstar{0}(\helig)\,\colfac{2}{\Gamma}\,,
\label{eq:OLrecUinit2l}
\ee
and, as discussed in the following subsections, we combine it with the two-loop numerator
in order to obtain a single helicity-summed object
\bea
\calU(q_1,q_2) &=& \sum_{\helig}\calU_0(\helig)\, \calN(q_1,q_2,\helig).
\label{eq:colborninterf2l}
\eea
We note that possible symmetry factors can be included in the
two-loop numerator $\calN$ or, alternatively, in the associated colour factor
$\calC_{2,\Gamma}$.

%

\subsection{Generic structure of a recursive two-loop algorithm}
\label{eq:genericalgo}

Thanks to its factorised structure, the colour and helicity summed two-loop
numerator~\refeq{eq:colborninterf2l} can be constructed 
through a numerical recursion of the form
\be
\hat{\calU}_n(q_1,q_2)=\hat{\calU}_{n-1}(q_1,q_2) \cdot \calK_{n}(q_1,q_2)\,\quad 
\label{eq:genericalg}
\ee
which starts with $\hat{\calU}_0=\idop$ and terminates, after a
certain number of steps $N_{\sss{r}}$, with
$\hat{\calU}_{N_{\sss{r}}}(q_1,q_2)=\calU(q_1,q_2)$.
We refer to the operations~\refeq{eq:genericalg} as dressing steps. 
The building blocks $\calK_n$ that are attached in each step 
are either the colour--Born interference $\calU_0$,
a connecting vertex, individual loop segments, or full chain numerators, \ie
\be
\calK_{n} \in \{\calU_0,\vertex{a},\segment{i}{k},\numc{i}\}\,.
\ee
The chain numerators $\numc{i}$ can in turn be constructed
through a similar recursion from
their loop segments. To this end we define the partially dressed chains
\bea
\numpc{i}{n}(q_i,\helipc{i}{n}) &=& \segment{i}{0}(q_i)\cdots\segment{i}{n}(q_i,\helisegment{i}{n}) 
\quad\text{and} \quad 
\numpc{i}{-1} =\idop\,,
\label{eq:def_pdc}
\eea
with helicities
\be
\helipc{i}{n} = \sum\limits_{a=1}^{n} \helisegment{i}{a}\,, \label{eq:heldef_pdc}
\ee
where $\helipc{i}{0}=0$ and $\helipc{i}{N_i-1}=\helic{i}$. 

In order to control the polynomial dependence on the loop
momenta in an efficient way, similarly as in the one-loop case, 
the dressing steps~\refeq{eq:genericalg} and all relevant building
blocks are implemented at the level of tensor coefficients. 

For instance, for the partially dressed chains~\refeq{eq:def_pdc} we use
the tensorial representation
\be
\numpc{i}{n}(\helipc{i}{n}) \,=\, 
\sum\limits_{r_i=0}^{n}\,
\numc{i}_{n;\mu_1 \ldots \mu_{r_i}}(\helipc{i}{n})\, q_i^{\mu_1}\ldots
q_i^{\mu_{r_i}}\,,
\ee
and the full chains
$\numc{i}(\helic{i}) = \numpc{i}{N_i}(\helic{i})$
are handled in a similar way. The individual loop segments in~\refeq{eq:OLrec}
have the same form as in
\eqref{eq:seg3point} and \eqref{eq:seg4point}. 
Three-point vertices $\vertex{0}$ and $\vertex{1}$ can be written as
\bea
\Big[\vertex{a}(q_1,q_2,q_3,\helisegment{V}{a})\Big]_{\indc{1}{a_1}\indc{2}{a_2}\indc{3}{a_3}}
&=& 
\Big[\hat{Y}_{a}\Big]_{\indc{1}{a_1}\indc{2}{a_2}\indc{3}{a_3}}\! 
+ \sum\limits_{i=1}^{3}\Big[\hat{Z}_{ia,\nu}\Big]_{\indc{1}{a_1}\indc{2}{a_2}\indc{3}{a_3}} q_i^\nu\,,
\label{eq:va3point}
\eea
where $a_i=0$ for $a=0$, $a_i=N_i-1$ for $a=1$, and $\helisegment{V}{a}=0$.
As observed above, triple vertices are helicity-independent, 
while the coefficients $\hat{Y}_{a}$ and $\hat{Z}_{ia,\nu}$ encode the 
linear dependence on the loop momenta.
Quartic vertices 
are independent of the loop momenta and
have the form
\bea
\Big[\vertex{a}(q_1,q_2,q_3,
\helisegment{V}{a})\Big]_{\indc{1}{a_1}\indc{2}{a_2}\indc{3}{a_3}}
&=& \Big[\hat{Y}_{a}^{\sigma}\Big]_{\indc{1}{a_1}\indc{2}{a_2}\indc{3}{a_3}}
\, w_{a\,\sigma}^{(V)}(\helisegment{V}{a})\,.
\label{eq:va4point}
\eea
The dependence of the two-loop numerator~\refeq{eq:colborninterf2l} on 
the two independent loop momenta $q_1,q_2$
is encoded in the tensorial representation
\bea 
\calU(q_1,q_2) &=&
\sum_{r=0}^{R_1} \sum_{s=0}^{R_2}\,
{\calU}_{\mu_1 \cdots \mu_{r}, \nu_1 \cdots \nu_{s}}
\,q_1^{\mu_1}\cdots q_1^{\mu_{r}}\,q_2^{\nu_1}\cdots q_2^{\nu_{s}}\,,
\label{eq:U2trep}
\eea 
and its contribution to the two-loop scattering probability density is
\bea 
\calW_{02,\Gamma} \,=\,
\frac{1}{\nhcs}\,
\sum_{r=0}^{R_1}\sum_{s=0}^{R_2} 
\,\mathrm{Re} \Big[\,
{\calU}_{\mu_1 \cdots \mu_{r}, \nu_1 \cdots \nu_{s}}
\,I^{\mu_1 \cdots \mu_{r}, \nu_1 \cdots \nu_{s}}\Big]\,,
\label{eq:W2d_tensordec}
\eea 
with the two-loop tensor integrals
\bea
I^{\mu_1 \cdots \mu_{r}, \nu_1 \cdots \nu_{s}} &=&
\int\!\rd\momq_1\!\int\!\rd\momq_2\,
\frac{q_1^{\mu_1}\cdots q_1^{\mu_{r}}\,q_2^{\nu_1}\cdots
q_2^{\nu_{s}}}{\denc{1}\,\denc{2}\,\dencexpl{3}{-\momq_1-\momq_2}}\,.  
\label{eq:A2d_tensorint}
\eea
The CPU efficiency and the memory footprint of 
the dressing recursion~\eqref{eq:genericalg} depend
in a critical way on the tensorial and helicity structure of the
various building blocks, which depend in turn on the order in which they are
attached to each other. 
In general, the most relevant aspects for the efficiency of the algorithm
are the number of independent tensor coefficients, and the structure of the Feynman rules
involved in the chain segments $\segment{i}{k}$ and in the connecting
vertices $\vertex{a}$.

At step $n$ of the construction, 
the structure of the first term on the rhs of~\eqref{eq:genericalg} is
\be
\hat{\calU}_{n-1}(q_1,q_2,\heli_{n-1}) 
\,=\, \sum\limits_{r=0}^{R_{1}}\sum\limits_{s=0}^{R_{2}} 
\Big[\hat{\calU}_{n-1}(\heli_{n-1})\Big]_{\mu_1\ldots\mu_{r},\nu_1\ldots\nu_{s}}^{\beta_1\ldots\beta_{N_l}} \, q_1^{\mu_1}\cdots q_1^{\mu_{r}} q_2^{\nu_1}\cdots q_2^{\nu_{s}}.
\label{eq:Uhattdec}
\ee
Here the number of independent tensor structures $q_1^{\mu_1}\ldots q_2^{\nu_s}$
grows exponentially with the rank in $q_1$ and $q_2$
as illustrated in \refta{tab:nitc},
and the tensorial rank increases during the construction.
In renormalisable theories, attaching a loop segment $\segment{i}{k}$ 
or a connecting vertex $\vertex{a}$ augments 
the rank in a loop momentum $q_i$ by either $0$ or $1$. 
In addition, the
number of tensor coefficients in~\refeq {eq:Uhattdec}
increases by a factor $4$ for each one of the 
open Lorentz/spinor indices $\beta_1,\dots, \beta_{N_l}$.
Such indices are associated with the various loop chains, and 
their number is typically \mbox{$N_l=0,2$ or $3$}
in the  phases of the construction that have a significant CPU cost.  
Finally, the number of tensor coefficients is also proportional to
the number of independent helicity configurations $\heli_{n-1}$ 
in \refeq{eq:Uhattdec}. 
Analogous considerations hold for the term $\calK_n$
on the rhs of~\eqref{eq:genericalg}.

\begin{table}[t]
\caption{Number of independent tensor structures 
$q_1^{\mu_1}\ldots q_1^{\mu_r}q_2^{\nu_1}\ldots q_2^{\nu_s}$ symmetrised in the $q_1$ and $q_2$ indices
up to maximal ranks $R_{1,2}$ in $q_{1,2}$, \ie for 
$r\leq R_1$ and $s \leq R_2$.
\label{tab:nitc}}
\begin{center}
\begin{tabular}{c|ccccc|}
{\normalsize ${}_{R_1}^{\quad R_2}$} & 0 & 1 & 2 & 3 \\ \hline
0 & 1   & 5   & 15   & 35   \\
1 & 5   & 25  & 75   & 175  \\
2 & 15  & 75  & 225  & 525  \\
3 & 35  & 175 & 525  & 1225 \\
4 & 70  & 350 & 1050 & 2450 \\
5 & 126 & 630 & 1890 & 4410 \\
\end{tabular}\end{center}
\end{table}

In general, we observe that the construction of a single chain
has the same complexity as for a one-loop chain, due to the dependence on a
single loop momentum and two open indices, while the steps involving the
two-loop vertices $\vertex{a}$ are much more expensive, since they involve
three indices and 
they typically
depend on two loop momenta.

Based on these qualitative considerations we have selected from
all possible algorithms of the form \eqref{eq:genericalg} the most promising
candidates.
The algorithm presented in the following section 
has been identified as the most efficient candidate 
based on a CPU cost simulation for a wide range of QED and QCD
Feynman diagrams. In this simulation
the CPU cost was approximated by the number of
numerical multiplications, which represent the most expensive operations.

It is worth mentioning that the algorithm presented in 
\refse{sec:twoloopalg_singledia} is roughly two orders of magnitude faster 
than a naive implementation, in which
all three chains are first constructed independently of each other
and are then contracted with the vertices 
$\vertex{0}$ and $\vertex{1}$. The inefficiency of this naive approach is due to the fact that the expensive $\vertex{0}$ and $\vertex{1}$ contractions are carried out when
the number of active helicities, Lorentz/spinor indices and tensor rank in the loop momenta are all maximal.

\subsection{Integrand of a single irreducible two-loop diagram}
\label{sec:twoloopalg_singledia}

In this section we present a dressing algorithm of type~\refeq{eq:genericalg}
for the construction of the integrand of a single
irreducible two-loop diagram of the form~\refeq{eq:A2d}
and its interference with the Born amplitude.
The organisation of the full set of irreducible two-loop diagrams is discussed
in~\refse{sec:twoloopalg_fullproc}.

As a starting point, given an irreducible two-loop diagram,
its chains and connecting vertices 
are ordered in a well-defined way that 
will
correspond to the sequence of dressing steps. 

\paragraph{Ordering rules:}
The ordering is implemented by assigning the labels $\vertex{0},\vertex{1}$ and
$\calCh{1},\calCh{2},\calCh{3}$ to specific vertices and chains
as follows.

\begin{itemize}
\item[I)] The three chains are ordered by their number of segments in such a way that $N_1 \geq N_2 \geq N_3$. In case of equality, the number of helicity configurations along the chain is used as a criterion. In case of equality there, the minimal external particle index on the chain is used, as defined in \eqref{eq:fullpartset}.
\item[II)] The role of $\vertex{0}$ and $\vertex{1}$ is assigned to the connecting vertices according to the following ordered set of criteria:
\begin{itemize}
\item[(i)] A single three-gluon vertex or a ghost-gluon-vertex with a rank increment in $q_1$ or $q_3$ becomes $\vertex{0}$.\footnote{Since ghosts are not allowed as external particles, ghost-gluon vertices always appear in pairs as $\vertex{0,1}$. Hence, this rule covers all cases
involving ghost-gluon-vertices.}
\item[(ii)] A single two-loop four-gluon vertex becomes $\vertex{1}$.
\item[(iii)] Use the external-particle indices, 
as defined in
\eqref{eq:fullpartset}, and the position of these particles along the
chains:
Particle $1$ is required to be closer to $\vertex{0}$, in case of equality particle $2$, etc.
\end{itemize}
\end{itemize}
These ordering rules 
fix the direction in which the various chains are constructed, 
namely from vertex $\vertex{0}$ to $\vertex{1}$ as depicted in~\eqref{eq:A2d}.
Moreover they determine the order and the way in which
the three chains and the two connecting vertices are 
constructed and attached to each other.
For this reason, the ordering rules affect the efficiency of the dressing algorithm in
a significant way.
The following algorithm has been 
optimised by testing 
a wide range of ordering rules,
using for example different hierarchies of the criteria in I) 
and/or interchanging the roles of the three chains,
e.g.~making
$\calCh{3}$ the chain with the most helicity configurations.  Furthermore
the vertex rules 
have been
optimised for each combination of QED and QCD vertices
as $\vertex{0,1}$.  

The reasoning behind the above ordering rules will be
explained 
together with the steps of the algorithm for which they are relevant. This algorithm consists of two main parts.

\subsubsection*{Part 1: Construction of chain $\calCh{3}$}

The shortest chain, i.e.~$\calC_3$,
is fully dressed through the recursion steps
\be
\numpc{3}{n}(q_3,\helipc{3}{n}) = \numpc{3}{n-1}(q_3,\helipc{3}{n-1})\cdot\segment{3}{n}(q_3,\helisegment{3}{n})
\qquad\text{for}\quad n=0,\ldots,N_3-1\;, \label{eq:recursion_chain3}
\ee
with the initial condition $\numpc{3}{-1}=\idop$. 
The partially dressed chain
numerator $\numpc{3}{n}$ depends on the helicity configuration of the 
already dressed segments,
\be
\helipc{3}{n} = \helipc{3}{n-1} + \helisegment{3}{n}\,,
\ee
and the number of such helicity states
grows in each recursion step~\eqref{eq:recursion_chain3} by a factor equal
to the number of helicity configurations $\helisegment{3}{n}$ of the
currently dressed segment $\segment{3}{n}$.
Similarly as in~\eqref{eq:OLrec}--\eqref{eq:OLrecfullcoeffa},
the above recursion is implemented in terms of tensor coefficients,
and the result
\be
\numc{3}(q_3,\helic{3})=\numpc{3}{N_3-1}(q_3,\helipc{3}{N_3-1})\,,
\ee 
where $\helic{3}=\helipc{3}{N_3-1}$,
is used as a building block in part~2 of the algorithm.

\subsubsection*{Part 2: Construction of the full diagram}

This part of the algorithm deals with the construction of
the colour- and helicity-summed interference of the 
two-loop integrand  with the full Born amplitude.
This is achieved through a dressing recursion of the
form~\eqref{eq:genericalg},
where all dressing operations are implemented at the 
level of tensor coefficients as 
described in~\refse{eq:genericalgo}. 

The starting point 
of the recursion
is the colour--Born 
interference
term
$\hat{U}_0=\calU_0$, as defined in \eqref{eq:OLrecUinit2l}, and the kernels
of the subsequent dressing steps are
\be
\calK_n \in \{
\segment{1}{0},\ldots,\segment{1}{N_1-1},
\vertex{1},\numc{3},
\vertex{0},
\segment{2}{0},\ldots,\segment{2}{N_2-1} \}\,,
\ee
\ie the full integrand is built by starting from the
chain $\calC_1$, which is subsequently connected 
to the vertex $\calV_1$
and the chain $\calC_3$. Finally, the vertex $\calV_0$ and the chain $\calC_2$
are attached.

\paragraph{Step 1 -- Construction of chain $\calCh{1}$:}

The goal of this step is the construction of the 
numerator of the
longest chain $\calCh{1}$ 
interfered with the  colour--Born factor $\calU_0$, defined in \eqref{eq:OLrecUinit2l} 
combined with the colour--Born interference
factor~\eqref{eq:OLrecUinit2l} 
and summed over the relevant helicities, \ie
\bea
\numi(q_1,\helicc{1}) &=&
\sum\limits_{\helic{1}} \calU_0(\helig)\, \numc{1}(q_1,\helic{1}).
\label{eq:dressedC1BornCol}
\eea
Here $\helic{1}$ and $\helig$ correspond, respectively,
to the helicity configurations of the full chain $\calC_1$
and of the whole process.
Upon summation over $\helic{1}$, the quantity on the lhs 
depends only on 
\be
\helicc{1}=\helig - \helic{1}\,,
\label{eq:heldef_cc_g}
\ee
which corresponds to the  helicity state of the still undressed part 
of the 
two-loop diagram. 
The quantity~\refeq{eq:dressedC1BornCol} is constructed 
by connecting the various segments 
of $\calC_1$ through a sequence of  
dressing steps
 \bea
\numpi{n}(q_1,\helipcc{1}{n})
&=&
\sum\limits_{\helisegment{1}{n}}
\numpi{n-1}(q_1,\helipcc{1}{n-1})\cdot\segment{1}{n}(q_1,\helisegment{1}{n})
\label{eq:dressingstepC1}
\eea
for $n=0,\ldots,N_1-1$.
In the $n$-th dressing step,
the helicity d.o.f.~$\helisegment{1}{n}$
of the $n$-th segment are summed 
on-the-fly with the technique of~\refse{sec:loopampU}.
As a result, $\numpi{n}$ depends only on the helicity state of the undressed
part of the two-loop diagram,
\be
\helipcc{1}{n}= \helig - \helipc{1}{n}\,,
\label{eq:heldef_pdcc_g}
\ee
where $\helipc{1}{n}$ 
is defined in~\refeq{eq:heldef_pdc} and 
corresponds to the dressed part of $\calC_1$. 
Each time that a new segment is attached,
the number of remaining helicity configurations 
decreases as  $\helipcc{1}{n}=\helipcc{1}{n-1}-\helisegment{1}{n}$.
The initial condition for the recursion~\refeq{eq:dressingstepC1} is
\bea
\numpi{-1}(q_1,\helipcc{1}{-1})&=&\calU_0(\heli)\,,
\eea
with
$\helipcc{1}{-1}=\heli$,
and the last step results into 
\be
\numpi{N_1-1}(q_1,\helipcc{1}{N_1-1})=\numi(q_1,\helicc{1})\,,
\ee
where $\helipcc{1}{N_i-1}{}=\helicc{1}$.

Since $\calCh{1}$ is the longest chain, a large number of helicities is already summed during this part of the
algorithm, 
and a large number of segments is constructed with dependence on a single loop momentum.

\paragraph{Step 2 -- Connecting the $\vertex{1}$ vertex and chain
$\calCh{3}$:}

In this 
twofold step the chain $\calCh{1}$ interfered with $\calU_0$
is connected to $\vertex{1}$ and to
the previously constructed numerator of $\calCh{3}$,
and $\calCh{1}$ interfered with $\calU_0$,
\bea
\Big[ \numinter{1}(q_1,q_3,\helic{2}+\helisegment{V}{0}) \Big]_{\indci{0}\indciii{0}\indcii{N_2}} &=& \sum\limits_{\helic{3}}\sum\limits_{\helisegment{V}{1}}
\Big[ \numi(q_1,\helicc{1})\Big]_{\indci{0}}^{\indci{N_1}}
\Big[ \numc{3}(q_3,\helic{3})\Big]_{\indciii{0}}^{\indciii{N_3}} \nonumber \\ & & \times
\Big[\vertex{1}(q_1,
q_2,
q_3,\helisegment{V}{1})\Big]_{\indci{N_1}\indcii{N_2}\indciii{N_3}}{}
\Bigg|_{q_2=-(q_1+q_3)}.
\qquad
\label{eq:c1dressing}
\eea
The helicity d.o.f.~of $\calCh{3}$ and $\vertex{1}$
are summed over on the fly, and the
result $\numinter{1}$ depends 
only on the helicity states $\helic{2}+\helisegment{V}{0}$
of the undressed parts
$\calCh{2}$ and $\vertex{0}$.
Note that
$\numinter{1}$ can involve a high number of coefficients due to three open indices 
$\indci{0}\indciii{0}\indcii{N_2}$ and two independent loop
momenta. 
Thus, for efficiency reasons,
when implementing~\eqref{eq:c1dressing}
it is useful to anticipate the dependence of $\vertex{0}$ on 
the indices $\indci{0}$ and $\indciii{0}$, 
and to ignore from the beginning all
$\indci{0} \indciii{0}$ combinations for which $\vertex{0}$ vanishes.

The step~\refeq{eq:c1dressing} 
is implemented using a tensorial representation
of the form~\refeq{eq:Uhattdec} with 
$q_1$ and $q_3$ as independent loop momenta. 

\paragraph{Step 3 -- Connecting the $\vertex{0}$ vertex:}

In this step $\vertex{0}$ is connected to the previously constructed
object~\refeq{eq:c1dressing}, which consists
of the numerators of $\calCh{1}$, $\vertex{1}$ and
$\calCh{3}$  interfered with $\calU_0$, 
    \bea
    \Big[ \numinter{10}(q_1,q_2,\helic{2}) \Big]^{\indcii{0}}_{\indcii{N_2}} &=& \sum_{\helisegment{V}{0}} \Bigg\{
    \Big[ \numinter{1}(q_1,q_3,\helic{2}+\helisegment{V}{0}) \Big]_{\indci{0}\indciii{0}\indcii{N_2}} \nonumber \\ & & \times
    \Big[\vertex{0}(q_1,q_2,q_3,
\helisegment{V}{0})\Big]^{\indci{0}\indcii{0}\indciii{0}}\Bigg\}_{q_3=
-(q_1+q_2)}{}\,.
\label{eq:Uonezerostep}
    \eea
Here the number of open indices is reduced to two.
For quartic vertices $\calV_0$ the 
associated helicity d.o.f.~are summed on the fly, and 
the output depends only on the helicity $\helic{2}$ 
of the still undressed  chain $\calC_2$.

Also this step is implemented using a tensorial representation, 
but now we switch to the independent loop momenta $q_1,q_2$
by replacing \mbox{$q_3= -(q_1+q_2)$}.
For a generic polynomial
\bea
A(q_1,q_3) &=&
\sum\limits_{r=0}^{r_1}\sum\limits_{s=0}^{r_3}
A_{\mu_1\ldots\mu_r,\nu_1\ldots\nu_s}
q_1^{\mu_1}\ldots q_1^{\mu_r} q_3^{\nu_1}\ldots q_3^{\nu_s}\,,
\label{eq:trep13}
\eea
with maximum ranks $r_{1,3}$ in $q_{1,3}$,
the replacement \mbox{$q_3= -(q_1+q_2)$} results into 
a polynomial
\bea
\tilde A(q_1,q_2) &=&
A(q_1,-q_1-q_2) \,=\,
\sum\limits_{r=0}^{R_1}\sum\limits_{s=0}^{R_2(r)}
\tilde A_{\mu_1\ldots\mu_r,\nu_1\ldots\nu_s}\,
q_1^{\mu_1}\ldots q_1^{\mu_r} q_2^{\nu_1}\ldots q_2^{\nu_s}\,,
\label{eq:Atildeq1q2}
\eea
with maximum ranks $R_1=r_1+r_3$ and $R_2(r)=0,\ldots,r_3$
in $q_1$ and $q_2$. 
In our implementation of~\refeq{eq:Uonezerostep}, 
by default 
this replacement is applied only upon 
contraction of 
$\numinter{1}$
with $\vertex{0}$ and summation over $\helisegment{V}{0}$.
In some cases it can be more efficient to apply \mbox{$q_3= -(q_1+q_2)$} to 
$\numinter{1}$ already at the end of step~2. In QCD this is the case 
if the vertex $\vertex{1}$ in step~2 is a triple gluon vertex or a ghost vertex 
that does not increase the rank in $q_3$.

For an efficient implementation of the entire algorithm 
it is important to note that---as explained in the following---the CPU cost of step~3 
is much lower as compared to step~2.
For example, let us consider step~3 for the case of a  
triple-gluon vertex,
\bea
\vertex{0}
^{\;\indci{0}\indcii{0}\indciii{0}}
\,=\,
g^{\;\indci{0}\indcii{0}}
(q_1-q_2)^{\;\indciii{0}}
+\mathrm{permutations}\,,
\label{eq:triplevertex}
\eea
where the prefactor with the coupling constant has been suppressed and 
can be restored as an overall factor 
at the end of the construction.
When this triple vertex is attached to 
the tensor 
$\calU_1^{(13)}$
via~\refeq{eq:Uonezerostep},
the various metric tensors in~\refeq{eq:triplevertex}  
give either rise to a summation over the
components of $\calU_1^{(13)}$  with $\indci{0}=\indciii{0}$, or to the
identification of the open index $\indcii{0}$ with one of the summed 
indices $\indci{0}, \indciii{0}$.
As for the loop momenta $q_i$ in~\refeq{eq:triplevertex}, their 
effect is simply to raise the tensor rank in $q_i$.
More explicitly, let us assume that 
$\calU_1^{(13)}$  is parametrised in terms of 
$q_1, q_3$ as in~\refeq{eq:trep13}.
In practice, exploiting the symmetry of tensor integrals 
wrt permutations of the indices associated with 
the individual loop momenta, such tensors can be 
implemented in the symmetrised form
\bea
A(q_1,q_3) &=&
\sum\limits_{n_0,\, \dots,\,m_3}
A_{n_0,\, \dots,\, n_3;\, m_0,\,\dots,\,m_3}
\prod_{\mu=0}^3\left(q_1^\mu\right)^{n_\mu}
\prod_{\nu=0}^3\left(q_3^\nu\right)^{m_\nu}\,,
\label{eq:symmtrep13}
\eea
where the tensor-coefficient indices $n_\mu$ and $m_\nu$ 
correspond to the powers 
in $q_1^\mu$ and $q_3^\nu$, respectively.
In this symmetrised representation, multiplying 
$A(q_1,q_3)$ by a component of $q_1$ or $q_3$ amounts to a trivial 
reshuffling of the tensor coefficients. For instance,
\bea
B(q_1,q_3) &=&
q_1^\beta \,A(q_1,q_3)
\eea
corresponds simply to  
\bea
B_{n_0,\,\dots,\,n_\beta,\,\dots,\,m_3} &=&
A_{n_0,\,\dots,\,n_\beta-1,\,\dots\,,m_3}\,. 
\eea
In summary, in case of a 
triple-gluon vertex $\calV_0$
step~3 does not involve any multiplication, but only 
sums and reshuffling operations, which are by far less 
time consuming.

Let us now consider step~2, where the 
vertex $\calV_1$ is connected. This operation can involve a very
large number of
multiplications, which can be estimated as follows.
The combination of the chains $\calC_1$ and $\calC_3$, \ie the tensors 
$\numi$ and 
$\numc{3}$
in~\refeq{eq:c1dressing}, 
requires $4^4=256$ multiplications 
due to the four open Lorentz/spinor indices
${\indci{0}}{\indci{N_1}}$ and ${\indciii{0}}{\indciii{N_3}}$.
Taking into account the number of helicity configurations
$\Nhel{i}$ and the number of tensor coefficients 
$\Ncoeff{i}$ of both chains, $\calC_i=\calC_1,\calC_3$, 
yields $256\times \Nhel{1}\Ncoeff{1}\Nhel{3}\Ncoeff{3}$
multiplications.
The remaining operations depend on the form of the Feynman rules for the 
connecting  vertex $\calV_1$ in~\refeq{eq:c1dressing}. Let us consider again a triple-gluon vertex,
\bea
{\vertex{1}}^{
\indci{N_1}
\indcii{N_2}
\indciii{N_3}
}
\,=\,
g^{\;\indci{N_1}\indcii{N_2}}
(k_1-k_2)^{\;\indciii{N_3}}
+\mathrm{permutations}\,,
\label{eq:triplevertexB}
\eea
with $k_i=q_i+p_{i,N_i-1}$, where  $p_{i,N_i-1}$ is the external
momentum of the last propagator along  $\calC_i$. 
Similarly as for step~3, the metric tensors in~\refeq{eq:triplevertexB}  
do not give rise to any additional multiplication.
The same holds for $q_i$ terms, while
the various $p_{i,N_i-1}$ terms can give rise to 
$16\time36\times 256\times \Nhel{1}\Ncoeff{1}\Nhel{3}\Ncoeff{3}$
additional multiplications.
In total, thanks to additional optimisations, our implementation 
of step~2 for a triple-gluon vertex  involves a number of multiplications 
slightly above $700\times \Nhel{1}\Ncoeff{1}\Nhel{3}\Ncoeff{3}$.
Given the potentially large number of helicity and 
tensorial components, this number can range from 
order $10^5$ to $10^7$ for nontrivial two-loop diagrams.

In general, step~2 is the most expensive operation of the entire algorithm,
and its exact cost depends on the nature of the 
connecting vertex $\calV_1$. 
For this reason, the CPU efficiency of the entire algorithm
depends in a critical way on the 
ordering rule II, which assigns the 
roles of $\vertex{1}$ and $\vertex{0}$
to the actual connecting vertices of 
two-loop diagrams. 
Regarding point (i) of the ordering rule~II, 
complex vertices are preferably chosen to be $\vertex{0}$.
In QCD the triple-gluon vertex is by far the most CPU expensive one, and as
a consequence it is only assigned the role of $\calV_1$ if both connecting vertices $\calV_{0,1}$ are of this type.
The rules for ghost--gluon vertices are guided by which rank of a loop momentum $q_i$ is increased, and hence at which point in the algorithm the replacement $q_3=-(q_1+q_2)$ is performed.
Regarding point (ii), a quartic
vertex is preferred to be inserted as $\vertex{1}$, since in this way the on-the-fly summation of the 
helicities $\helisegment{V}{1}{}$ of its external subtree can be performed at an earlier stage.

Finally we note that when 
both $\vertex{0}$ and $\vertex{1}$ are fermion--gauge-boson vertices, 
whose tensorial structure consists of
$\gamma$-matrices containing 
only a few non-vanishing components, the most efficient implementation available in our framework is to  combine $\vertex{1}$ and $\vertex{0}$ into a single tensor $\Gamma(\vertex{1},\vertex{0})$
and 
to merge steps~2 and~3 in a single operation,
    \bea
    \Big[\numinter{10}(q_1,q_2,\helic{2})\Big]^{\indcii{0}}_{\indcii{N_2}} &=& \sum_{\helic{3}}\Bigg\{
    \Big[ \numi(q_1,\helicc{1})\Big]_{\indci{0}}^{\indci{N_1}}
    \Big[ \numc{3}(q_3,\helic{3})\Big]_{\indciii{0}}^{\indciii{N_3}} \nonumber \\ & & \times
    \Big[\Gamma(\vertex{1},\vertex{0})\Big]_{\indci{N_1}\indcii{N_2}\indciii{N_3}}^{\indci{0}\indcii{0}\indciii{0}}
    \Bigg\}_{q_3= -(q_1+q_2)}{},
    \eea
where the index sums run only over the non-zero components.
This is particularly efficient in pure QED calculations, where all vertices are proportional to $\gamma^\mu$.

\paragraph{Step 4 -- Connecting $\calC_2$:} 

The construction of the 
two-loop numerator is completed by attaching the chain $\calC_2$.
The various segments of $\calC_2$ are 
connected through a sequence of dressing steps
 \bea
\label{eq:chaintwodress}
 \numpd{n}(q_1,q_2,\helipccloc{2}{n}) &=& \sum\limits_{\helisegment{2}{n}}
  \numpd{n-1}(q_1,q_2,\helipccloc{2}{n-1}) \cdot \segment{2}{n}(q_2,\helisegment{2}{n})
 \eea
for $n=0,\ldots,N_2-1$. 
At step $n$ the helicity d.o.f.~$\helisegment{2}{n}$
of the $n$-th segment are summed on the fly, 
and the result depends only on the helicities of the
still undressed segments,
\be
\helipccloc{2}{n} \,=\, 
\sum\limits_{a=n+1}^{N_2-1} \helisegment{2}{a}{}
\,=\,
\helipccloc{2}{n-1}- \helisegment{2}{n}\,.
\label{eq:heldef_pdcc_l}
\ee
The initial condition for the recursion is given by the outcome of 
step~3, \ie 
\be
 \numpd{-1}(q_1,q_2,\helipccloc{2}{-1}) \,=\,
\numinter{10}(q_1,q_2,\helic{2})\,,
 \ee 
where $\helipccloc{2}{-1} = \helic{2}$ corresponds to the
helicity configuration of the full chain $\calC_2$.

The dressing steps~\refeq{eq:chaintwodress}
are similar to the ones for the construction of $\calCh{1}$, since the segments $\segment{2}{n}$ depend on a single loop momentum $q_2$. 
However, the dressing of chain $\calC_2$ features 
an higher complexity 
due to the dependence of $\numpd{n}$ on two loop momenta, which leads to a larger number of tensor components.

During the last step the loop is closed by contracting the index $\indcii{N_2}$ between vertex $\vertex{1}$ and chain $\calCh{2}$,
 \bea
 \calU(q_1,q_2) &=& 
\Tr\left[\numpd{N_2-1}(q_1,q_2,\helipccloc{2}{N_2-1})\right] 
\nonumber\\ &=& 
 \sum\limits_{\helisegment{2}{N_2-1}}
 \Big[ \numpd{N_2-2}(q_1,q_2,\helipccloc{2}{N_2-2}) \Big]^{\indcii{N_2-1}}_{\indcii{N_2}}
 \cdot 
 \Big[ \segment{2}{N_2-1}(q_2,\helisegment{2}{N_2-1})\Big]^{\indcii{N_2}}_{\indcii{N_2-1}},
 \eea
where $\helipccloc{2}{N_2-1}=0$ and
$\helipccloc{2}{N_2-2}=\helisegment{2}{N_2-1}$.

\begin{figure}[t!]
\begin{center}
\includegraphics[width=0.38\textwidth]{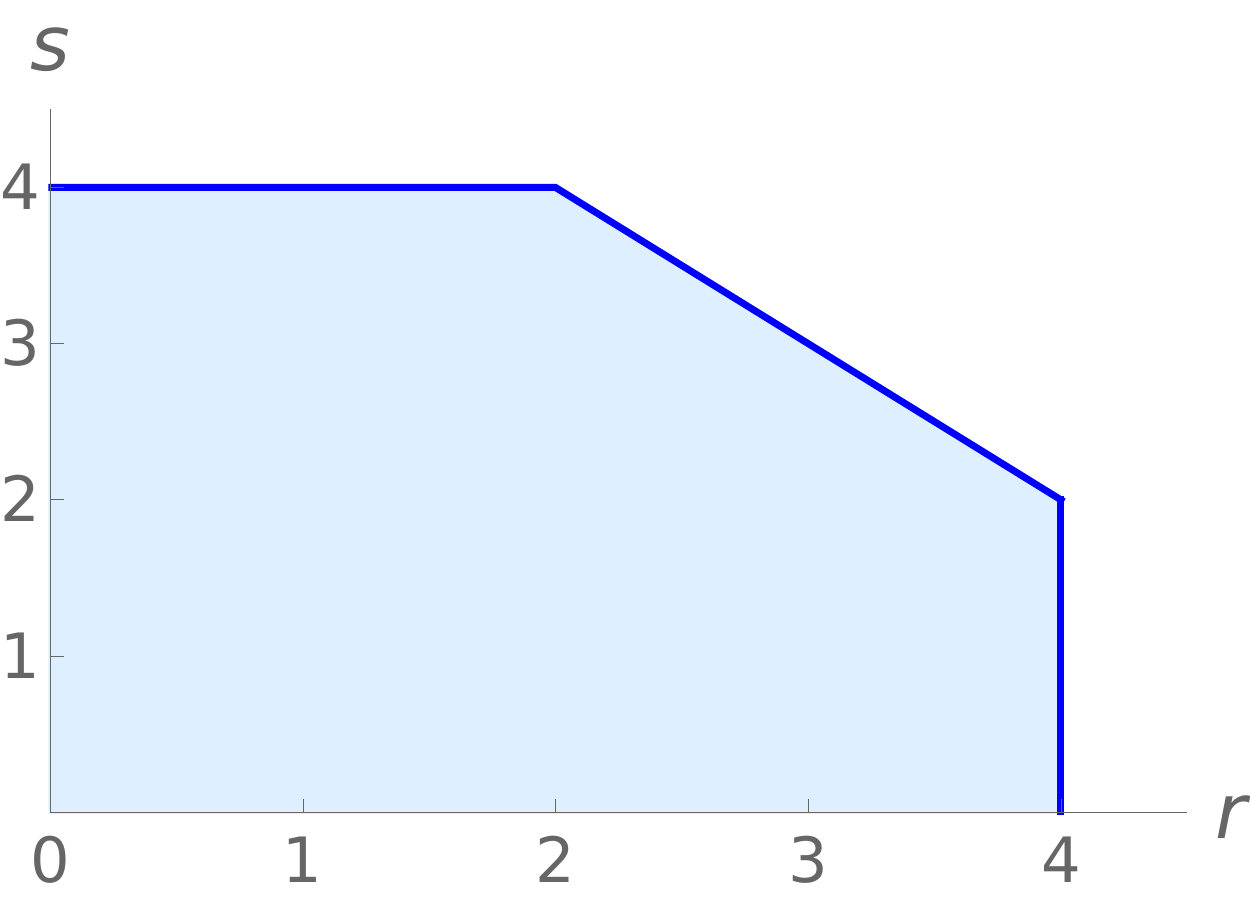} \vspace{-4mm}
 \end{center}
\caption{Example of the interdependence of the tensorial ranks. The 
coloured area shows the ranks $(r,s)$ in $(q_1,q_2)$ 
that result from the \mbox{$q_3=-(q_1 + q_2)$} replacement
applied to a polynomial with ranks $r_1,r_2,r_3\le 2$ 
in $q_1, q_2, q_3$.
In this case the maximum rank in $q_2$ obeys
$R_2(r)\le \min(4,6-r)$.
Note that the excluded region with 
$r,s\le 4$ and $ s+r>6$ corresponds to the highest-rank configurations,
which involve a large number of tensor coefficients.
}
 \label{fig:example_rankdep} 
\end{figure}

\paragraph{Efficient implementation of steps 2--4:}
The steps~2 to~4 involve polynomials
of the form~\refeq{eq:trep13} and~\refeq{eq:Atildeq1q2},
with a large number of tensor coefficients.
For their efficient implementation, besides using symmetrised 
representations like~\refeq{eq:symmtrep13}, 
we systematically exploit the correlation of 
tensorial ranks as explained in the following.
Let us consider polynomials in the two independent loop momenta $q_1$ and $q_2$,
which consist of various monomials of rank $(r,s)$ in $(q_1,q_2)$.
As a result of the particular form of the $q_1, q_2$
dependence of the involved Feynman rules, the maximum ranks 
$R_1=\max(r)$ and $R_2=\max(s)$
are not independent.
For instance, attaching a triple-gluon vertex 
can increase the rank $(r,s)$ 
of a monomial to $(r+1,s)$ or $(r,s+1)$ but not to 
$(r+1, s+1)$.
In general, as a result of such correlations, 
for the subset of tensor coefficients with fixed rank $r$ in $q_1$,
the maximum rank in $q_2$ depends on $r$, 
\ie $R_2 = R_2(r)$, and vice versa.
As discussed in~\refeq{eq:Atildeq1q2}, this 
kind of correlation arises also from 
the \mbox{$q_3=-(q_1 + q_2)$} replacement.
An example of the corresponding $R_2(r)$ dependence is shown 
in~\reffi{fig:example_rankdep}.

In our implementation, for each single step of the algorithm 
the values of $R_1$ and $R_2(r)$ are determined
based on the actual $q_1,q_2$ dependence of the 
involved Feynman rules, and all operations are
restricted to the tensor components
with $r\le R_1$ and $s\le R_2(r)$.
In particular, we systematically avoid memory allocation for irrelevant tensor
components.
As a result the overall memory usage is typically reduced by more than a factor two as compared to an implementation with
independent $R_1$ and $R_2$.

\subsection{Two-loop tensor coefficients for 
scattering amplitudes} \label{sec:twoloopalg_fullproc}

For a set of Feynman diagrams, in particular for all irreducible two-loop diagrams of a full scattering
amplitude, the algorithm is structured in such a way that the overall number of dressing steps is minimised. This 
is
achieved by identifying all equivalent dressing steps 
that are
required in multiple diagrams and performing such steps only once.
To this end, all diagrams are ordered 
as
discussed in \refse{sec:twoloopalg_singledia}, and  
the steps described above for a single diagram are performed for the full set of diagrams in the following way. 
\paragraph{Part 1:}
 The chains $\calCh{3}$ of all diagrams are grouped according to their number of segments $N_3$, and the various groups are dressed 
starting from the lowest $N_3$, \ie the chains with 
only a
propagator segment $\segment{3}{0}$. 
All fully or partially dressed chains that have already been constructed are
stored and can be recycled whenever they appear as part of another chain. 
Since the various chains are ordered such that $N_3 \leq N_2 \leq N_1$, all
chains $\calCh{3}$ involve only a few segments,\footnote{For 
a $2 \to 3$ process $\calCh{3}$ does not have more than two segments.} 
which require a rather small number of dressing equations.
 
 \paragraph{Part 2:}
 For the main algorithm, where the full two-loop amplitudes are constructed, we employ an extension of the on-the-fly diagram merging procedure, first introduced
in~\cite{Buccioni:2017yxi} for one-loop amplitudes. This approach exploits the factorisation of the numerators of Feynman diagrams into segments. Consider for instance two diagrams with the same 
loop
topology, \ie the same scalar propagator denominators, which 
have the same undressed segments
after a certain number $n$ of recursion steps,
e.g.~
\bea
\calU_A &=&
\numpi{A,n}\cdot \segment{1}{n+1}\cdots\segment{1}{N_1-1}\cdot
\numc{3}\cdot
\vertex{1} \cdot
\vertex{0} \cdot \segment{2}{0}\cdots\segment{2}{N_2-1}\,,\nonumber \\
\calU_B &=&
\numpi{B,n} \cdot\segment{1}{n+1}\cdots\segment{1}{N_1-1}\cdot
\numc{3}\cdot
\vertex{1} \cdot
\vertex{0}  \cdot \segment{2}{0}\cdots\segment{2}{N_2-1} \,.
\eea
Here $\numpi{A,n}$ and
$\numpi{B,n}$ can differ due, for instance, 
to a different external subtree in the \mbox{$n$-th} 
segment of the chain $\calC_1$.
Since ultimately the sum of all two-loop 
diagrams interfered with the full Born amplitude is needed, the partial numerators $\numpi{A,n}$ and $\numpi{B,n}$
can be 
summed, and all subsequent steps, here 
from segment $\segment{1}{n+1}$ to segment $\segment{2}{N_2-1}$,
are performed only once, 
\ie
\bea
\calU_A + \calU_B &=&
\lb\numpi{A,n} + \numpi{B,n}\rb\cdot\segment{1}{n+1}\cdots\segment{1}{N_1-1}\cdot
\numc{3}\cdot
\vertex{1} \cdot
\vertex{0}  \cdot 
\segment{2}{0}\cdots\segment{2}{N_2-1} {}
\,.
\eea
To this end, the full set of two-loop diagrams is
constructed \textit{segment by segment}, \ie each of the
steps~1
to~4 in 
\refse{sec:twoloopalg_singledia}
is performed on all diagrams before continuing with the next step. For example, the first dressing step $\numpi{-1}\segment{1}{0}$ is performed for all diagrams before all diagrams are dressed with segment $\segment{1}{1}$. After each dressing step along chain $\calCh{1}$ and $\calCh{2}$
the algorithm recognises and performs all possible merging 
operations
before proceeding with the next segment. After the final dressing and merging steps all diagrams with the same topology, and hence corresponding to a single type of tensor integral, are summed into a single object.

The last criterion in ordering rule I, 
see~\refse{sec:twoloopalg_singledia},
namely the ordering of chains by external particle indices in case of $N_i=N_j$ and the same number of helicities,
is designed in such a way that
on-the-fly merging opportunities during the construction 
of sets of diagrams are maximised. 
In general,
merging steps can be performed for any number of diagrams, independently of their ranks.
The overall CPU efficiency benefit of the on-the-fly merging is $10$ to $20\%$ for most processes.

Alternatively, to the \textit{segment-by-segment} approach, the full set of recursion steps 1 to 4 can be performed \textit{diagram by diagram}, without merging. This makes it possible to free the memory of previously computed partially dressed diagrams, thereby minimising memory usage. Since memory is usually not the bottleneck of the calculation, we chose the {segment-by-segment} approach as default.

\section{Technical performance of the two-loop algorithm}
\label{eq:performance}
The two-loop algorithms presented
in~\refses{sec:twoloopred}{sec:twoloopirred} have been implemented in the
\OpenLoops{} framework in a fully automated way.  The present implementation
supports QED and QCD corrections to any SM process,
and the full generality of our method allows for extensions 
to any other model.

This section is devoted to the validation and the technical performance of
our implementation.  In particular, we discuss the numerical stability, CPU
efficiency and memory footprint for a variety of $2\to 2$ and $2\to 3$
processes at two loops in QED and QCD.
Since the construction of reducible two-loop diagrams 
is largely based on well-established one-loop techniques, 
in the following we restrict ourselves to the contributions stemming 
from irreducible two-loop diagrams, which are 
constructed with the algorithm of~\refse{sec:twoloopirred}.

\subsection{Validation with pseudotree test} 

For the validation of the two-loop algorithm we have 
implemented a so-called
\textit{pseudotree test}, which
was employed to validate every new
numerical routine, e.g.~all possible two-loop vertices $\vertex{0,1}$, in
single diagrams, as well as the full algorithm for a wide range of
processes.

 \begin{figure}[t!]
\begin{center}
\includegraphics[width=0.5\textwidth]{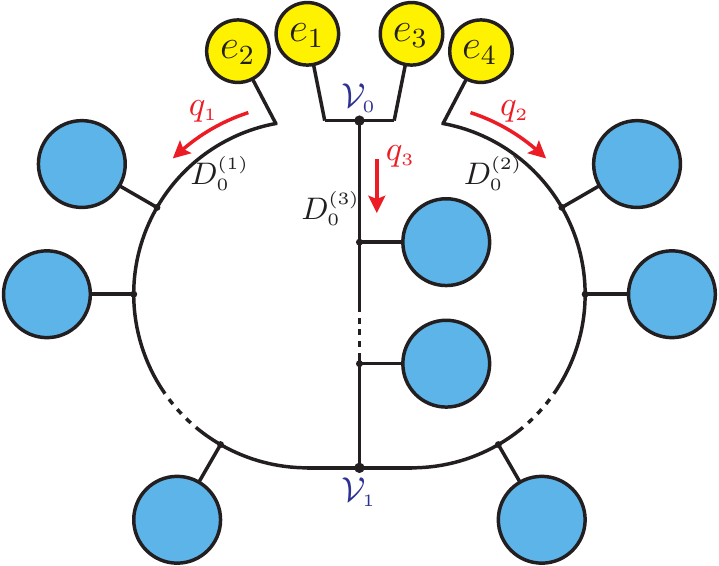} 
 \end{center}
 \caption{Insertion of pseudo wave functions into an irreducible two-loop
diagram.}
 \label{fig:pseudotreeID}
\end{figure}

The idea of the pseudotree test is to cut-open $L$-loop Feynman diagrams 
by ``cutting'' $L$ loop lines, 
such as to obtain tree diagrams that can be 
evaluated with well established tree-level algorithms.
In the case of irreducible two-loop diagrams,
as illustrated in~\reffi{fig:pseudotreeID},
we cut the loop lines associated with $D^{(1)}_0$ and  $D^{(2)}_0$ 
on the side of the vertex $\vertex{0}$. 
The resulting open Lorentz/spinor indices are contracted with four pseudo
wave functions $e_1,\ldots,e_4$, which are filled with random numbers.
The loop momenta $q_1$ and $q_2$ are 
also fixed at random values,
and the incoming external momenta associated with $e_1,e_2,e_3,e_4$ are 
$-q_1,q_1,-q_2,q_2$.

In the pseudotree test the cut-open diagrams of
\reffi{fig:pseudotreeID}  
are constructed in two different ways,
as detailed in the following.

First, the colour- and helicity-summed interference of the 
cut-open two-loop diagram with the Born amplitude
is computed using
the well-tested tree-level algorithm~(t)
in \OpenLoops. This calculation is carried out
at fixed external momenta $q_1, q_2$, and 
is implemented in double precision (DP) and quadruple
precision (QP), where the latter is intended as a benchmark.  The Born
two-loop interferences computed in this way are denoted
$\widehat{\calW}_{02}^{(\rm{t,DP})}$ and $\widehat{\calW}_{02}^{(\rm{t,QP})}$, 
and
\be
\widehat{\calW}_{02} \,=\, \sum\limits_\Gamma \widehat{\calW}_{02,\Gamma}\,, \label{eq:integrandsum}
\ee
where the sum runs 
over the full set of irreducible two-loop diagrams $\Gamma$ of the process 
at hand, and $\widehat{\calW}_{02}$ denotes 
the loop integrand of the ${\calW}_{02,\Gamma}$ defined in \eqref{eq:M2WvirtBL4Gamma}.

Second, we compute the pseudo trees with the two-loop algorithm (2L), constructing the original
two-loop diagram with pseudo segments $e_1^{\gamma_1}e_2^{\gamma_2}$ and $e_3^{\gamma_3}e_4^{\gamma_4}$ inserted between $\vertex{0}$ and the cut propagators $D^{(1)}_0$ and $D^{(2)}_0$.
The resulting loop numerator $U(q_1,q_2)$, 
defined in~\refeq{eq:colborninterf2l},
is obtained in the tensorial representation~\refeq{eq:U2trep},
which encodes  the full $q_1,q_2$ dependence in the form of tensor coefficients. 
Including all denominators, we obtain
\be
{\widehat{\calW}_{02,\Gamma}^{(\rm{2L})}}=\f{U(q_1,q_2)}{\denc{1}\,\denc{2}\,\denc{3}} \Big|_{q_3=-(q_1+q_2)},
\label{eq:twoloopptt}
\ee
which is evaluated at fixed $q_1,q_2$ 
and compared against the result obtained with the tree algorithm.
The Born two-loop interference computed in this way is denoted $\widehat{\calW}_{02,\Gamma}^{(\rm{2L,DP})}$ 
and $\widehat{\calW}_{02,\Gamma}^{(\rm{2L,QP})}$, for a single diagram in DP and QP respectively, 
and for a full process we compute $\widehat{\calW}_{02}^{(\rm{2L,DP})}$ and $\widehat{\calW}_{02}^{(\rm{2L,QP})}$ 
according to \eqref{eq:integrandsum}.

To validate our algorithm with the pseudotree test
we have determined the relative numerical uncertainty 
\be
\calA_{\rm{p}}^{(\rm{t})} := \log_{10}\left(\f{|\widehat{\calW}_{02}^{(\rm{t,p})}-\widehat{\calW}_{02}^{(\rm{2L,p})}|}{\text{Min}(|\widehat{\calW}_{02}^{(\rm{t,p})}|,|\widehat{\calW}_{02}^{(\rm{2L,p})}|)}\right),
\label{eq:pttest}
\ee
where p=DP,QP. This test was applied to
a wide range of single Feynman diagrams and full SM processes, 
considering two-loop QED or QCD corrections.

\begin{table}[b]
\caption{
Processes considered for the 
validation of the two-loop algorithm 
and for the assessment of its CPU efficiency 
and memory footprint (see
\refses{sec:cpu_eff}{se:memory}).
The massive and massless states that enter 
closed fermion loops are listed in the
third and fourth columns.
Light quarks are implemented as one or two
first-generation fields. In processes with a single internal quark field $u$,
light-fermion loops are multiplied with the number of light quarks $N_l$ in order to restore all active massless quark flavours.
If the doublet $(u,d)$ is used, light-fermion loops are multiplied with the number of light generations $N_l/2$.}
\label{tab:procdetails}
\begin{center}
 \begin{tabular}{c|c|c|c|c}
  corrections & process type & massless fermions & massive fermions & process \\ \hline 
  QED & $2 \to 2$ & $e$ & $-$ & $e^+e^-\to e^+e^-$ \\ \cline{2-5}
      & $2 \to 3$ & $e$ & $-$ & $e^+e^-\to e^+e^-\gamma$ \\
  \hline
  QCD & $2 \to 2$ & $u$ & $-$ & $gg\to u\bar{u}$ \\
      &  & $u,d$ & $-$ & $d\bar{d}\to u\bar{u}$ \\
      &  & $u$ & $-$ & $gg\to gg$ \\
      &  & $u$ & $t$ & $u\bar{u}\to t\bar{t}g$ \\
      &  & $u$ & $t$ & $gg\to t\bar{t}$ \\
      &  & $u$ & $t$ & $gg\to t\bar{t}g$ \\
  \cline{2-5}
      & $2 \to 3$ & $u,d$ & $-$ & $d\bar{d}\to u\bar{u}g$ \\
      &  & $u$ & $-$ & $gg\to ggg$ \\

      &  & $u,d$ & $-$ & $u\bar{d}\to W^+gg$ \\
      &  & $u,d$ & $-$ & $u\bar{u}\to W^+W^-g$ \\
      &  & $u$ & $t$ & $u\bar{u}\to t\bar{t}H$ \\
      &  & $u$ & $t$ & $gg\to t\bar{t}H$ \\
 \end{tabular}\end{center}
\end{table}
In particular, we have investigated the 
$2\to 2$ and $2\to 3$ processes
listed in~\refta{tab:procdetails}, which 
involve all possible incarnations of the 
QED and QCD Feynman rules in the various steps of the algorithm of~\refse{sec:twoloopirred}.
The double-precision comparison $\calA_{\rm{DP}}^{(\rm{t})}$ 
showed excellent agreement for all tested diagrams and processes with 
typical accuracy at the level of $10^{-15}$.

\subsection{Numerical stability}
 \begin{figure}[t]
    \begin{center}
     \begin{tabular}{cc}
   \quad $gg \to t \bar{t}$ & \quad $d\bar{d} \to u\bar{u}g$ \\[-1mm]
    \includegraphics[trim = 1mm 1mm 1mm 10mm, clip, height=52mm]{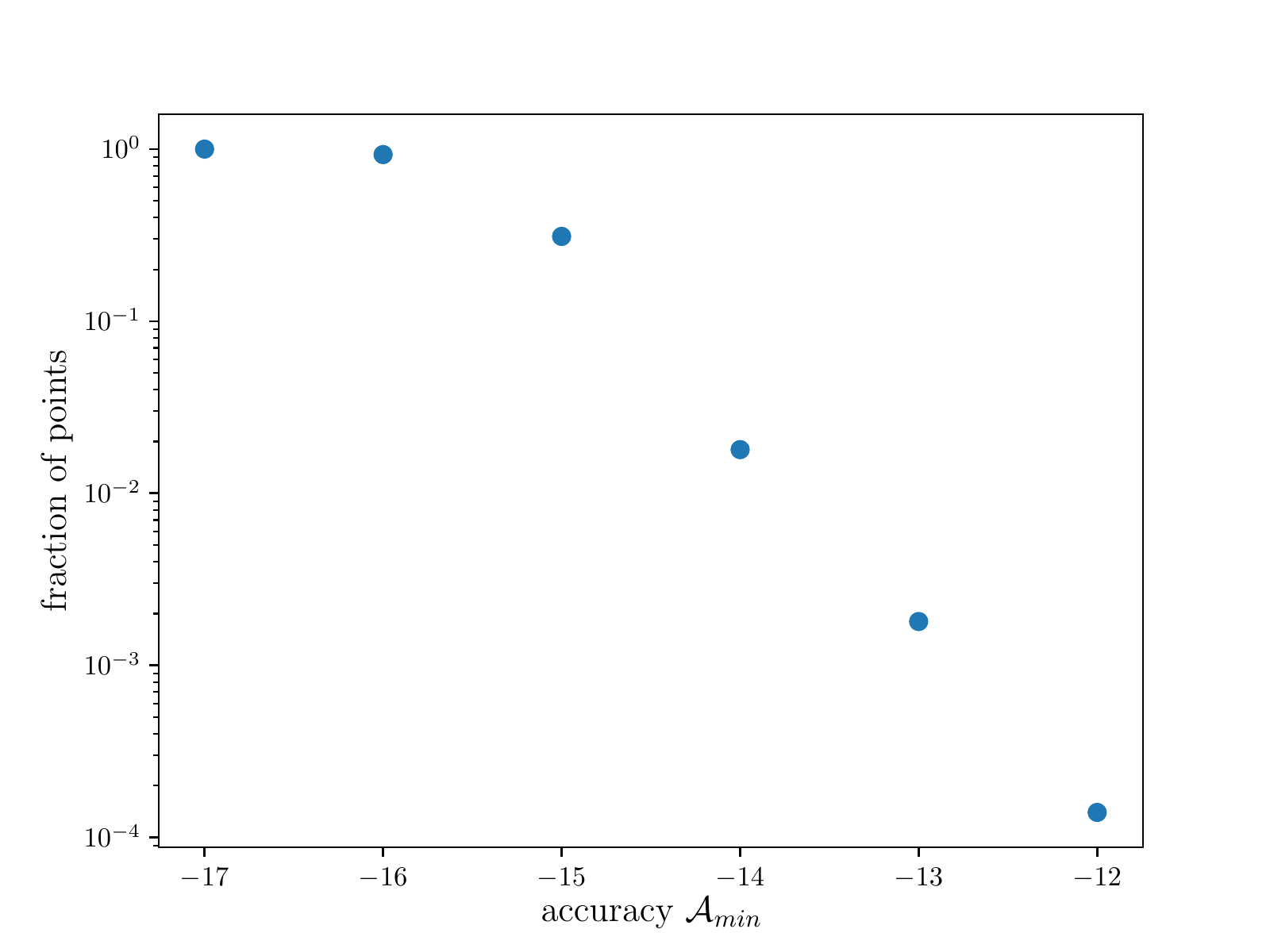} 
    &
    \includegraphics[trim = 1mm 1mm 1mm 10mm, clip, height=52mm]{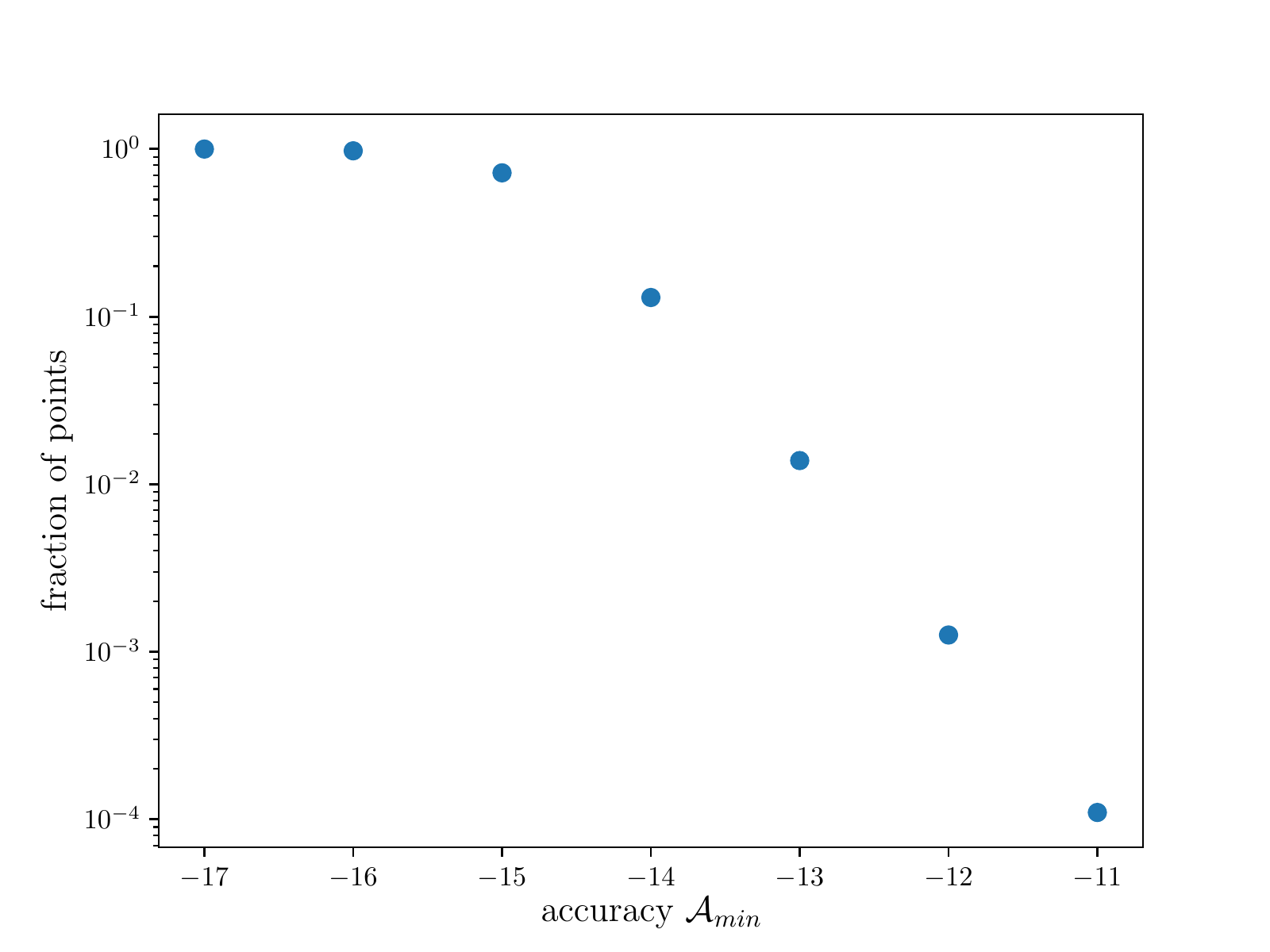} 
    \end{tabular}    
    \end{center}

    \caption{Numerical stability of the two-loop algorithm  
for irreducible diagrams in pseudotree mode.
The plots show the 
fraction of phase-space points with 
a double-precision instability 
$\calA_{\rm{DP}}>A_{\rm{min}}$ as a function of 
$A_{\rm{min}}$. The distributions correspond to 
$gg\to t \bar t $ (left) and $d\bar{d} \to u\bar{u}g$ (right)
samples with $10^5$ events.
} 
\label{fig:numstabPT}
\end{figure} 

In order to demonstrate the numerical stability of the two-loop algorithm
we have assessed the relative deviation between 
numerical evaluations of~\refeq{eq:twoloopptt} in DP and QP,
\be
\calA_{\rm{DP}} \,=\,
\log_{10}\left(\f{|\widehat{\calW}_{02}^{(\rm{2L,DP})}-\widehat{\calW}_{02}^{(\rm{2L,QP})}|}{\text{Min}(|\widehat{\calW}_{02}^{(\rm{2L,DP})}|,|\widehat{\calW}_{02}^{(\rm{2L,QP})}|)}\right)\,,
\label{eq:twoloopnumacc}
\ee
To this end, we have first 
established the QP result $\widehat{\calW}_{02}^{(\rm{2L,QP})}$ 
as a benchmark by confirming that the 
pseudotree test~\refeq{eq:pttest} in QP yields a typical
accuracy at the level of $10^{-30}$ and always much better than $10^{-17}$, which is the limit accuracy in
DP.

The numerical accuracy~\refeq{eq:twoloopnumacc}
was tested for selected processes using samples 
of $10^5$ uniformly distributed phase-space points.
Results for a $2\to 2$ and a $2\to 3$ process are reported in 
\reffi{fig:numstabPT}.
In both cases the bulk of the points has a
relative uncertainty of $10^{-16}$ to $10^{-14}$, and the upper bounds
for the relative uncertainties are of order $10^{-12}$ and $10^{-11}$ for 
the $2\to 2$ and the $2\to 3$ process, respectively.

In conclusion, the two-loop algorithm for the construction
of tensor coefficients is fully implemented and validated for QED and QCD corrections. Its numerical stability is similarly good as for the tree-level and one-loop \OpenLoops{} algorithm.
This is an important prerequisite for a full two-loop calculation, where the 
dominant numerical instabilities are expected to arise from 
the reduction of tensor integrals, 
and should not be further enhanced by the 
calculation of tensor coefficients.

\subsection{CPU efficiency} \label{sec:cpu_eff}

In this section we assess the CPU cost
of the two-loop algorithm of~\refse{sec:twoloopirred}
and---in order to judge its impact in the context of realistic applications---we  
compare it to the cost of 
the related real--virtual contributions
to a full NNLO calculation. 

\begin{figure}[t!]
   \begin{center}
     \includegraphics[width=\textwidth]{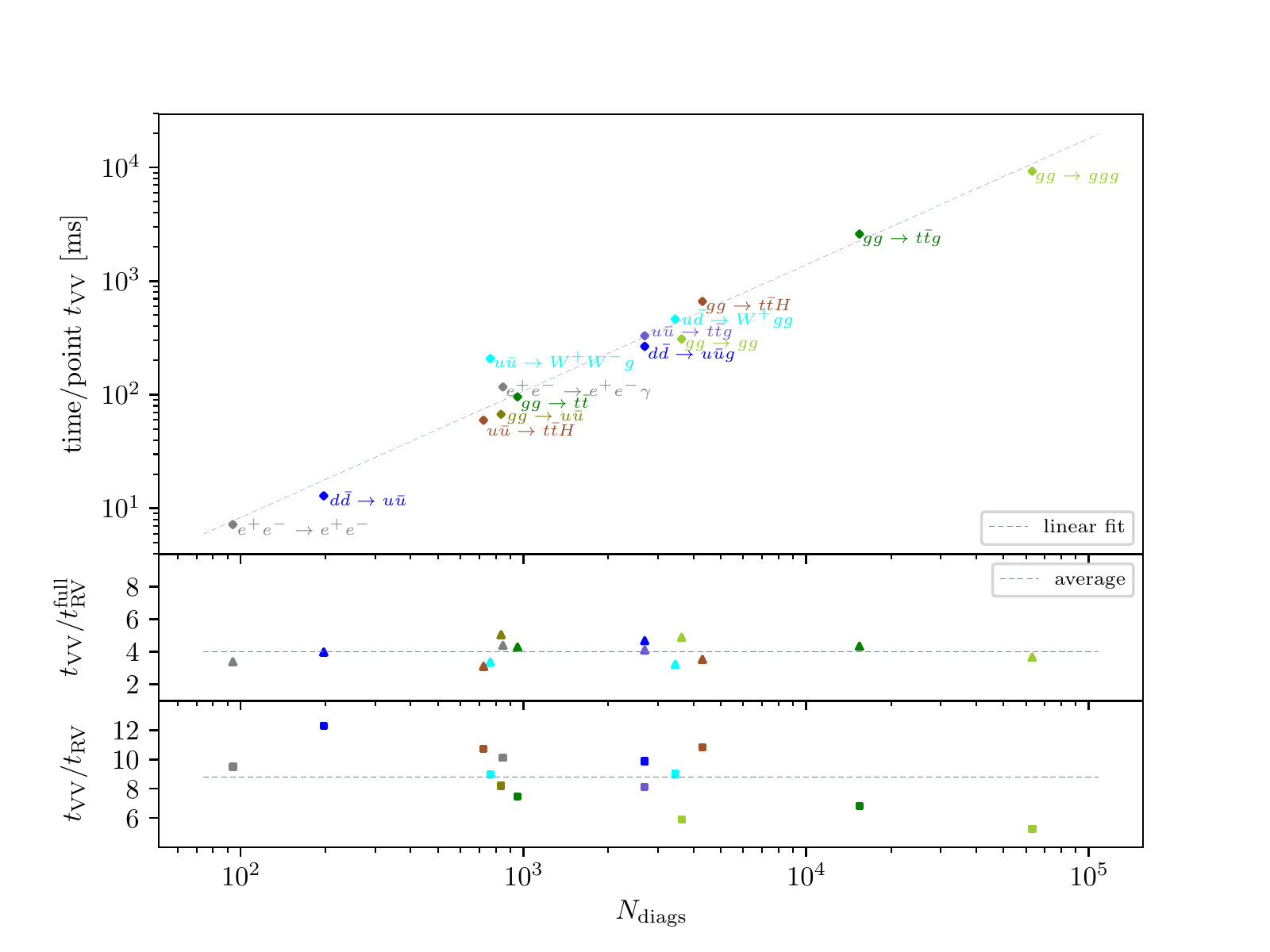} 
   \end{center}
   \caption{Runtimes per phase-space point for the calculation of the
two-loop tensor coefficients on a single core
Intel(R)~Core(TM)~i7-6600U\,@\,2.6\,GHz with 16GB RAM.  The timings
correspond to the QED and QCD processes in \refta{tab:procdetails} and are
plotted versus the number $\Ndiags$ of irreducible two-loop diagrams.  The
upper frame displays the absolute two-loop timings $\tVV$ in ms.  
The middle frame shows the ratio $\tVV/\tRVfull$, where $\tRVfull$ is the time for computing the 
full one-loop
probability density $\calW_{01}^{(1)}$
(including tensor integrals) for a related process
with one extra gluon or photon
which enters the real--virtual part of the same NNLO calculation. 
The lower frame shows the ratio $\tVV/\tRV$, where $\tRV$ is the cost of
computing the one-loop tensor coefficients of $\calW_{01}^{(1)}$
without tensor integrals.
Colour and helicity sums are included throughout, 
and the latter are always carried out on-the-fly.
All timings have been
taken in DP using the GNU Fortran 7.5.0 compiler and averaging over 1000
points.  
}
\label{fig:efficiency_plot} 
 \end{figure}

The CPU cost of the algorithm of~\refse{sec:twoloopirred}
is influenced by several aspects, such as 
the number of Feynman diagrams, the presence of massive
propagators, the number of external helicity configurations (2 for fermions,
3 for $W$ and 1 for $H$ bosons) and the complexity of the vertices, in particular
the connecting vertices $\vertex{0,1}$, as well as the maximum tensor ranks. 
The overall impact of these different aspects is illustrated in
\reffi{fig:efficiency_plot}, where we present the total CPU cost 
for a 
wide range of processes with two-loop QED and QCD corrections. Again we
consider the $2\to 2$ and $2\to 3$ processes listed
in~\refta{tab:procdetails}, which involve all relevant building blocks of
our algorithm in different combinations.  
Colour and helicity sums are included throughout, and
 all two-loop timings in \reffi{fig:efficiency_plot} correspond to the 
\textit{segment-by-segment} approach.

In the upper frame of \reffi{fig:efficiency_plot}
we present the virtual--virtual timings per phase-space point, $\tVV$,
for the construction of the two-loop tensor
coefficients. 
Plotting these timings versus the number of 
irreducible two-loop diagrams $\Ndiags$ we observe 
clusters of processes of similar number of diagrams,
which also have similar timings.  The simplest $2
\rightarrow 2 $ processes, such as $e^+ e^- \to e^+ e^-$, take around
10\,ms/point.  Next there is a group of more complex $2 \rightarrow 2 $ and
simple $2 \rightarrow 3$ processes, which require 
about $100$\,ms/point, the
fastest one being $u \bar{u} \to t \bar{t} H$ at $65$\,ms/point.  The next group
consists of more complex $2 \rightarrow 3$ processes, the fastest one being
$d \bar{d} \to u \bar{u} g$ at $261$\,ms/point, as well as the most expensive
$2 \rightarrow 2 $ process, $gg \to gg$.  The most time consuming processes
we studied are $gg \to t \bar{t}H$, which takes $2.5$\,s/point, and $gg \to
ggg$, which takes $9.2$\,s/point.  The timings plotted in
\reffi{fig:efficiency_plot} are 
presented in \refapp{app:cpu_eff}.

In general, we observe that the two-loop 
runtimes grow only linearly with the number of diagrams,
\ie the ratio $\tVV/\Ndiags$
is approximately a process-independent constant. 
On the employed CPU we find
\bea
\tVV
\approx
\Ndiags\times 
150\,\mu\mathrm{s}\,.
\eea
We also observe that 
adding an extra external 
gluon (photon) to a process in QCD (QED) increases the number of diagrams as well as the computation time 
$\tVV$ by approximately one order of magnitude,
e.g.~in $e^+ e^- \to e^+ e^- (+\gamma)$ and $gg \to t \bar{t} (+g)$. 
These process-independent scaling features are
comparable to the behaviour of one-loop calculations in \OpenLoops.

In the lower frames of \reffi{fig:efficiency_plot} 
we show the ratio of virtual--virtual and real--virtual 
timings for partonic subprocesses that contribute to the same NNLO calculation. 
More precisely, 
the virtual--virtual time $\tVV$, which corresponds to the 
two-loop tensor coefficients for 
a given $2\to\npart$ process, 
is compared to the real--virtual time
$\tRV$ corresponding to the  
one-loop correction
to the associated $2\to\npart+1$ process 
with one extra gluon or photon.
Helicity d.o.f~ are always summed one-the-fly, both at one and two
loops.

The real--virtual timings $\tRVfull$
in the middle frame
correspond to the cost of the
full real--virtual density $\calW_{01}^{(1)}$ in \eqref{eq:M2WwithX1}, 
\ie  in $\tRVfull$ we
include also the cost of 
the required one-loop tensor integrals.\footnote{In this case the full integral
reduction and evaluation is performed after the construction of the tensor
coefficients with \Collier~\cite{Denner:2016kdg}.
} 
In this way we gain an idea of the share of the two-loop tensor coefficient construction in a full NNLO calculation.
As shown in \reffi{fig:efficiency_plot},
in spite of the fact that $\tVV$
and $\tRVfull$ vary by more than three orders of magnitude,
their ratio amounts to 
\bea
\frac{\tVV}{\tRVfull} 
&\approx& 4\pm 1\,,
\label{eq:VVtoRV ratio}
\eea
and is approximately constant.
In other words, independently of the type of process and 
its particle multiplicity, the cost of two-loop tensor coefficients per phase-space point is only a few times higher as compared to 
the related real--virtual amplitudes.

The lower frame of \reffi{fig:efficiency_plot} 
shows the same ratio for real--virtual timings $\tRV$ that do not include 
the contribution of tensor integrals.
This provides a more direct comparison of the cost of one-loop and 
two-loop NNLO building blocks at the level of tensor coefficients.
For this ratio we find
\bea
\frac{\tVV}{\tRV} 
&\approx& 9\pm 3\,,
\label{eq:VVtoRV ratio coeff}
\eea
and again we observe an approximately process-independent behaviour.
This is related to the fact 
that the number of two-loop diagrams 
for a given process usually lies between the number of one-loop diagrams for the 
corresponding process with one and two extra
partons (gluon or photons). We also note that
the total tensor rank in the loop momenta is the same at two-loop level and
at one-loop level with two extra 
partons, and comparing the cost of the corresponding 
tensor coefficients we find that 
$\tVV$ is a factor 3 to 8 less expensive wrt the one-loop 
coefficients for processes with 
two extra partons (see \refapp{app:cpu_eff}).  
Considering the much
higher complexity of two-loop diagrams as compared to one-loop diagrams,
these are very promising findings.

Finally, let us compare the efficiency  of the 
\textit{diagram-by-diagram} and \textit{segment-by-segment} 
approaches 
discussed in \refse{sec:twoloopalg_fullproc}.
The former
 uses less memory and tends to be slower, while the latter
requires more memory and tends to be faster thanks to the
on-the-fly merging of diagrams.  
For simple processes the two approaches 
feature similar timings, their ratio being $1 \pm 0.15$, and in most cases the \textit{segment-by-segment} approach being faster. For more complex processes, however, the 
\textit{segment-by-segment} approach outperforms the \textit{diagram-by-diagram} approach by a factor $1.15$ to $1.35$, with the highest ratio of $1.35$ for $gg \to ggg$. This behaviour is expected, since the efficiency gain due the on-the-fly merging is larger for 
more complex processes with a higher number of Feynman diagrams.
As the default option for our two-loop algorithm the \textit{segment-by-segment} approach is the clear choice.

In conclusion, 
for the same number of phase-space points  
the CPU cost of two-loop tensor coefficients 
is of the same order as 
the cost of the related real--virtual
parts of a NNLO calculation.
However, the number
of required points for the integration of 
the two-loop contributions is
typically much lower.
In the light of these observations, the presented
two-loop timings 
are quite promising. In practice we expect that 
the cost of two-loop tensor coefficients 
should play a subleading role 
in the CPU budget of complete $2\to 3$ NNLO
calculations.

\subsection{Memory Usage}
\label{se:memory}

\begin{table} 
\caption{Memory required by the full set of tensor coefficients 
for the processes of \refta{tab:procdetails}.
The two central columns correspond to the construction of 
two-loop coefficients in the \textit{segment-by-segment} and \textit{diagram-by-diagram} approaches
(see~\refse{sec:twoloopalg_fullproc}).
The last two columns report the memory footprint
of the one-loop coefficients
for the  real--virtual contributions to the same 
NNLO calculation, as well as the memory footprint of the full real--virtual calculation, including the reduction and evaluation of the 
tensor integrals with {\sc Collier}. For these the public \OpenLoopstwo{} program with on-the-fly helicity summation was used.
} 

\label{tab:memory_usage_2l}
\begin{center}
\begin{tabular}{|l|c|c||c|c|}
\hline
& \multicolumn{2}{c||}{virtual--virtual memory [MB]}
& \multicolumn{2}{c|}{real--virtual [MB]} \\ \hline
hard process    & segment-by-segment     & diagram-by-diagram 
& coefficients & full 
\\\hline\hline
$e^+ e^- \rightarrow e^+ e^-$ & 18  & 8  & 6  & 23 \\ \hline
$e^+ e^- \rightarrow e^+ e^- \gamma $ & 154  & 25  & 22 & 54 \\ \hline
$gg \rightarrow u \bar{u} $ & 75  & 31  & 10 & 26 \\ \hline
$gg \rightarrow t\bar{t}$ & 94  & 35  & 15 & 34 \\ \hline
$gg \rightarrow t\bar{t}g$ & 2000  & 441 & 152 & 213 \\ \hline
$u \bar{d} \rightarrow W^+gg$ & 563  & 143  & 54 & 90 \\ \hline
$u\bar{u} \rightarrow W^+ W^- g$ & 264  & 67 & 36 & 67 \\ \hline
$u\bar{u} \rightarrow t\bar{t}H$ & 82  & 28  & 14 & 40 \\ \hline
$gg \rightarrow t\bar{t}H$ & 604  & 145  & 50  & 90 \\ \hline
$u\bar{u} \rightarrow t\bar{t}g$ & 323  & 83  & 41 & 74 \\ \hline
$gg \rightarrow gg$ & 271  & 94  & 41 & 55 \\ \hline
$d\bar{d} \rightarrow u \bar{u} $ & 18  & 10  & 9 & 20 \\ \hline
$d\bar{d} \rightarrow u\bar{u}g$ & 288  & 85  & 39 & 68 \\ \hline
$gg \rightarrow ggg$ & 6299  & 1597  & 623 & 683 \\ \hline
\end{tabular}
\end{center}
\end{table}

The construction of two-loop tensor coefficients 
with the  algorithm of~\refse{sec:twoloopirred}
can require an
important amount of memory.  This is quantified in~\refta{tab:memory_usage_2l} for
the full set of processes listed in~\refta{tab:procdetails}.
In addition to the memory footprint of the \textit{segment-by-segment} and
\textit{diagram-by-diagram} approaches at two loops, for comparison
we also report the corresponding memory usage for real--virtual contributions, 
\ie for the one-loop corrections to the related  processes
with one extra parton.
At two loops only the memory for the 
storage of tensor coefficients is considered, while at one loop we present both the memory consumption for the construction of 
the tensor coefficients and for the full calculation.

We observe that the 
memory requirements for the two-loop tensor coefficients
in the \textit{diagram-by-diagram} and
\textit{segment-by-segment} approaches are, respectively,
a factor $1$--$3$ and $2$--$13$ higher wrt the
real--virtual case. 
Including the one-loop tensor integrals, the \textit{diagram-by-diagram} and
\textit{segment-by-segment} require, respectively,
a factor $0.3$--$2.3$ and $0.8$--$9.4$ of the memory for the full real--virtual calculation.
The \textit{segment-by-segment} approach 
uses $2$ to $6$ times more memory than the 
\textit{diagram-by-diagram} approach for the construction of the tensor coefficients.
Thus the
latter may be preferable when 
memory consumption plays a critical role.  
However, this is never the case for the considered processes,
and both two-loop approaches are well within the scope of current computing
environments.

\section{Conclusion}\label{sec:conclusion}

We have presented a new and fully general algorithm for the
efficient construction of two-loop integrands.  To this end we have
designed and implemented a generalisation of the open-loops method to two loops.
In this approach, the polynomial dependence of two-loop integrand numerators
on the loop momenta $q_1, q_2$ is encoded in a set of tensor coefficients
that are constructed through a numerical recursion.
The interference with the Born amplitude as well as helicity and colour sums
are systematically included, and the resulting tensor coefficients are ready
to be combined with the relevant two-loop tensor integrals to form 
scattering probability densities at two loops.

For reducible two-loop diagrams, which 
factorise into one-loop subdiagrams,
we have presented an algorithm that exploits and extends various aspects of the
pre-existing \OpenLoops{} program at one loop, while for irreducible
two-loop diagrams we have developed a largely new and more sophisticated
algorithm, which represents the main novelty of this paper.

Irreducible two-loop diagrams involve 
three loop chains that depend, respectively,  on the loop momenta
$q_1$, $q_2$, $q_3=-q_1-q_2$, and are attached to each other through two 
connecting vertices.
Each loop chain corresponds to a sequence of
so-called loop segments, each consisting of a loop propagator 
attached to one or two external subtrees through a vertex.
The numerator of an irreducible two-loop diagram is 
constructed through a numerical recursion
where the various loop segments and connecting 
vertices are attached to each other one after the other.
The single steps of the recursion correspond to 
process-independent operations that depend 
only on the Feynman rules of
the model at hand, and the 
entire recursion is implemented in terms of tensor coefficients.
Due to
the large number of tensorial structures and helicity degrees of freedom,
the most expensive individual steps can require of the order of $10^6$
multiplications per two-loop diagram.
Thanks to a systematic comparison of several algorithmic options and a
detailed cost analysis we have identified an efficient algorithm, which 
turned out to outperform naive approaches by two orders of magnitude.
This high efficiency is achieved through an optimal ordering of the construction steps,
the factorisation of colour structures, the on-the-fly summation of helicity
degrees of freedom, and various other tricks.

The algorithm has been implemented in a fully automated way in
the \OpenLoops{} framework, and this first implementation 
supports two-loop QED and QCD corrections to 
any scattering process within the Standard Model.
Its correctness and numerical stability 
have been verified by means of a so-called  
pseudo-tree test in combination with quadruple-precision benchmarks.
In order to assess the efficiency of the new algorithm we have 
studied a set of 
$2\to 2$ and $2\to 3$ processes with  
numbers of two-loop diagrams ranging from order 
$10^2$ to $10^5$ per process.
We have observed that the cost of the two-loop tensor coefficients 
per phase-space point for 
an entire process is approximately 
proportional to the number of (irreducible) two-loop diagrams. 
On a single Intel~i7 core 
this cost is around 150\,$\mu$s per two-loop diagram, 
and for typical $2\to 3$ partonic processes at two loops in QCD
it is of the order of one second.
Finally we have shown that
the cost of the two-loop tensor coefficients 
that enter the virtual--virtual part of a 
NNLO calculation 
is comparable to the one of the
one-loop ingredients that enter
the corresponding real--virtual part.

In conclusion, the presented algorithm provides a key 
building block for the construction of an automated generator of 
scattering amplitudes at two loops, and the 
observed technical performance guarantees its applicability to 
arbitrary $2\to 2$ and $2\to 3$ processes.

\subsection*{Acknowledgments}
We would like to thank F.~Buccioni for useful discussions.
This research was supported by the Swiss National Science Foundation (SNSF) 
under the SNSF Ambizione grant PZ00P2-179877. The work of S.P.~was 
supported through contract BSCGI0-157722.

\newpage
\appendix
\section{CPU efficiency measurements} \label{app:cpu_eff}

In \refta{tab:timings} we report the explicit timings
that have been discussed in~\refse{sec:cpu_eff} and summarised 
in~\reffi{fig:efficiency_plot}.
The first column corresponds to the full list of 
hard scattering processes defined in~\refta{tab:procdetails}.
The second and third columns indicate, respectively, the number of 
irreducible two-loop diagrams, $\Ndiags$,
and the CPU time $\tVV$ per phase space point for
the construction of two-loop tensor coefficients, 
as defined in \refse{sec:cpu_eff}.
This time $\tVV$ corresponds the default variant of the 
two-loop
algorithm, \ie the \textit{segment-by-segment} approach 
(see \refse{sec:twoloopalg_fullproc}). The 
fourth column shows the ratio of $\tVV$
to the  corresponding time $\tVV^{\mathrm{dbd}}$ required by the
\textit{diagram-by-diagram} approach.
The next two columns give the ratios of $\tVV$ to the timings for the
one-loop tensor coefficients ($\tRV$) or full one-loop amplitude
including tensor integrals ($\tRVfull$)
for the corresponding process with one
additional parton (gluons in QCD and photons in QED).
The last two columns give the ratios of $\tVV$ and the timings for the
one-loop tensor coefficients ($\tRRV$) or full one-loop calculation
($\tRRVfull$) for the corresponding process with two extra partons.
All time measurements have been carried out as detailed in the caption
of~\reffi{fig:efficiency_plot}.

\begin{table}[t]
\caption{Virtual--virtual and real--virtual timings for the $2\to 2$ and $2\to 3$ processes 
discussed in \refse{sec:cpu_eff}.
\label{tab:timings}}
\begin{center}
\begin{tabular}{|p{2.7cm}|p{1cm}|p{1.4cm}|p{1.2cm}|p{1.2cm}|p{1.2cm}|p{1.2cm}| p{1.2cm}|}
\hline
process & $\Ndiags$ & $\tVV$ [ms] &    $\frac{\tVV}{\tVV^{\mathrm{dbd}}}$ 
&    $\frac{\tVV}{\tRV}$ &    $\frac{\tVV}{\tRVfull}$ &    $\frac{\tVV}{\tRRV}$ &    $\frac{\tVV}{\tRRVfull}$ \\ \hline\hline
$e^+e^-\to e^+e^-$ &    94 &   7.18  &   1.15&  10.1&   3.47&   0.633&   0.280\\ \hline
$e^+e^-\to e^+e^-\gamma$ &   848 & 114  & 1.10   &  10.0 &  4.44 &  0.578 & 0.267  \\ \hline
$gg \to u \bar{u} $ &   835 &  66.3  &   0.994&   8.67&   5.35&   0.309&   0.200\\ \hline
$d\bar{d} \to u \bar{u} $ &   197 &  12.7  &   1.04&  12.9&   4.08&   0.471&   0.231\\ \hline
$d\bar{d} \to u\bar{u}g$ &  2690 & 261  & 0.940 &9.71   &  4.76 &  0.339 &  0.203 \\ \hline
$gg \to gg$ &  3633 & 315  &   0.845&   5.86&   5.01&   0.185&   0.129\\ \hline
$gg \to ggg$ & 63105 & 9250  &  0.738  & 5.26   &3.68    & -   & -   \\ \hline
$u\bar{u} \to t\bar{t}g$ &  2690 & 309  &   0.964&   7.46&   3.97&   0.259&   0.165\\ \hline
$gg \to t\bar{t}$ &   955 &  91.9  &  1.04 & 7.10 & 4.14 & 0.251 & 0.158  \\ \hline
$gg \to t\bar{t}g$ & 15462 & 2541  &   0.957 & 6.93  & 4.37 &-&-\\ \hline
$u \bar{d}\to W^+gg$ &  3450 & 455  &   0.956&   8.96&   3.25&   0.334&   0.123\\ \hline
$u\bar{u} \to W^+ W^- g$ &   765 & 194  &   0.979&   8.68&   3.40&   0.362&   0.115\\ \hline
$u\bar{u} \to t\bar{t}H$ &   724 &  64.7  &   1.03&  11.8&   3.49&   0.408&   0.165\\ \hline
$gg\to t\bar{t}H$ &  4304 & 636  &   0.923&  10.9&   3.38&   0.361&   0.151\\ \hline
\end{tabular}
\end{center}
\end{table}

\newpage
\bibliographystyle{JHEP}
\bibliography{OL_2l_literature}

\end{document}